\RequirePackage{fix-cm}
\documentclass[smallextended]{svjour3}       
\smartqed  
\usepackage{graphicx}
\usepackage{multirow}
\usepackage{numprint}
\usepackage{textcomp}
\usepackage{booktabs}
\usepackage{amsmath}
\usepackage{amssymb}
\usepackage{textcomp}
\usepackage{dirtytalk}
\usepackage{float}
\usepackage{xcolor}
\usepackage{verbatim}
\usepackage{enumerate} 
\usepackage{xcolor}
\usepackage{colortbl}
\usepackage{breakcites}
\begin{document}

\title{An Efficient Approach for Reviewing Security-Related Aspects in Agile Requirements Specifications of Web Applications
}


\author{Hugo Villamizar \and Marcos Kalinowski \and Alessandro Garcia \and Daniel Mendez 
}


\institute{Hugo Villamizar 
            \email{hvillamizar@inf.puc-rio.br}    
            \and
            Marcos Kalinowski
            \email{kalinowski@inf.puc-rio.br}
            \and
            Alessandro Garcia 
            \email{afgarcia@inf.puc-rio.br}
            \and
            Daniel Mendez
            \email{daniel.mendez@bth.se}
}

\date{Received: date / Accepted: date}

\maketitle

\begin{abstract}

Defects in requirements specifications can have severe consequences during the software development lifecycle. Some of them may result in poor product quality and/or time and budget overruns due to incorrect or missing quality characteristics, such as security. This characteristic requires special attention in web applications because they have become a target for manipulating sensible data. Several concerns make security difficult to deal with. For instance, security requirements are often misunderstood and improperly specified due to lack of security expertise and emphasis on security during early stages of software development. This often leads to unspecified or ill-defined security-related aspects. These concerns become even more challenging in agile contexts, where lightweight documentation is typically produced. To tackle this problem, we designed an approach for reviewing security-related aspects in agile requirements specifications of web applications. Our proposal considers user stories and security specifications as inputs and relates those user stories to security properties via Natural Language Processing. Based on the related security properties, our approach identifies high-level security requirements from the Open Web Application Security Project (OWASP) to be verified, and generates a reading technique to support reviewers in detecting defects. We evaluate our approach via three experiment trials conducted with 56 novice software engineers, measuring effectiveness, efficiency, usefulness, and ease of use. We compare our approach against using: (1) the OWASP high-level security requirements, and (2) a perspective-based approach as proposed in contemporary state of the art. The results strengthen our confidence that using our approach has a positive impact (with large effect size) on the performance of inspectors in terms of effectiveness and efficiency.

\end{abstract}

\section{Introduction} \label{sec:introduction}

Requirements Engineering (RE) is an inherently complex part of software engineering. Given its complexity, defects such as ambiguities, inconsistencies and incomplete requirements may arise. These defects have been reported by practitioners to be causing problems in software projects, such as poor product quality and time and budget overruns ~\cite{fernandez2017naming}. Moreover, the costs for correcting these RE-related problems increases throughout the software development lifecycle \cite{boehm2005software}. These additional costs reinforce the importance of identifying such defects at early stages. 

The rapidly changing business environments in which many companies operate are challenging traditional RE approaches. This gave rise to agile methods for RE. Agile RE relies on lightweight documentation and face-to-face collaborations between customers and developers~\cite{cao2008agile}. Yet, agility does not necessarily compensate for problems that have been reported for plan-driven software process models (e.g., RUP, V-Model XT, Waterfall), such as moving targets and communication flaws ~\cite{fernandez2017naming}. It can even make those problems more explicit if a critical prerequisite for successful RE are not met: human-intensive exchange, collaboration, and trust~\cite{fernandez2015naming}. In other words, agile RE has helped to address some RE problems, but it has also hindered others, such as under-specified requirements that are too abstract~\cite{fernandez2017naming}. 

Security is an essential Non-Functional Requirement (NFR) that requires special attention of any software system that manages valuable data and processes, among others, due to business needs to protect restricted user information. Much of sensitive information is hosted on the internet, making web applications an often favoured target. In today's software systems, security vulnerabilities are increasing~\cite{deepa2016securing} and ensuring security controls is becoming more difficult. Defects in security requirements can lead to important vulnerabilities that impact core functionality, leading to loss of reputation, financial penalties, and even legal consequences. 
For instance, in September 2016, the internet giant Yahoo announced it had been the victim of the biggest data breach in history~\cite{FoxBusiness}. The hackers stole personal information of 500 million users. The attack was performed via phishing emails with a link. Once it was clicked, malware was downloaded to the network. Thus, bad actors gained access to the user database. This can be seen as a breach not covered by the security requirements. Not surprisingly, security is considered as a relevant NFR for the 21st century~\cite{penzenstadler2014safety}. 

For that reason, specifying and verifying security requirements is crucial to ensure software product quality. Nevertheless, security requirements are often misunderstood and improperly specified due to lack of security expertise and emphasis on security during early stages of software development~\cite{mellado2010systematic}. While software requirements inspections represent a promising approach to effectively verify security requirements, security expertise is essential, but often lacking in software engineers~\cite{carver2006finding}. Hence, software engineers could benefit from specific reading techniques to support the verification of security aspects during requirements inspections.

The picture is even more challenging in agile contexts. Most agile teams do not have a security expert on board. Therefore Product Owners (POs) and developers are, at best, responsible for identifying and prioritizing security requirements~\cite{daneva2018security}. However, POs as well as developers, often lack security knowledge~\cite{terpstra2017agile}. This may results in software which fails to properly deal with security. According to Eberlein~\cite{eberlein2002agile}, there is a need for agile methods to include techniques that make it possible to identify NFRs early. There is also a need to describe them in such a way that they may be analyzed early, thus reducing the likelihood of costly rework~\cite{mcgraw2006software}. Alsaqaf et al.~\cite{alsaqaf2017quality} conducted a literature review on engineering NFRs in agile projects that cover security concerns. They reported 12 challenges such as the inability of user stories, the most used artifact in agile RE, to document quality requirements, the product owner’s lack of knowledge, the dependence on the product owner as the single point to collect the requirements, and the delay in the validation of the requirements. That is why several recent secondary studies acknowledge the urgent need for methods to systematically engineer security requirements in agile projects~\cite{alsaqaf2017quality, villamizar2018systematic}. In that direction, Villamizar et al.~\cite{villamizar2018systematic} identified that most of the studies dealing with security in agile contexts, lack empirical evaluation and research on requirements verification. Such activities should be conducted to assure those agile requirements specifications are correct, consistent, unambiguous, and complete. This means appropriately covering basic security-related aspects, such as input validation, unauthorized access, assignment of administrative privileges and denial-of-service attacks.

Although these secondary studies have reported several challenges in literature, the number of solution proposals to address these concerns is limited. Existing approaches (e.g.,~\cite{carver2006finding}~\cite{elberzhager2009software}~\cite{peine2008security}~\cite{riaz2014hidden}) employ inspection techniques to verifying security or identify security goals from textual documents. These proposals are not focused on agile, but this does not mean that they cannot address these kinds of requirements. For instance, Carver et al.~\cite{carver2006finding} propose a perspective-based approach to identify security defects in general requirements. In this case, the authors do not explicitly mention considering their approach in an agile context, but that does not exclude it to cover textual requirements or agile specifications. To the best of our knowledge, only one study explicitly proposes a methodology to address security verification activities in agile software development~\cite{domah2015nerv}. The study proposes a lightweight methodology to address NFRs early in agile software development processes. Activities such as elicitation, reasoning and validation are considered within the methodology in an effort to maintain agility while at the same time attempting to improve the quality of software developed with agile processes. 

Given the contemporary state of reported evidence, we took a step forward to address the existing literature gap concerning security requirements verification in agile contexts. We proposed and evaluated an approach for reviewing security-related aspects in agile requirements specifications of web applications, previously presented in~\cite{villamizar2019approach}. We decided to focus on web applications given that they have become a target for accessing or extracting sensible data. Results of our initial experimental evaluation, comparing our approach against using the complete list of OWASP high-level security requirements, indicated that using our approach has a positive impact on the performance of inspectors in terms of effectiveness and efficiency.

In this paper, we extend our previous study~\cite{villamizar2019approach} in two ways: First, we provide further details on the approach and its initial evaluation, thus facilitating its adoption. Second, we conducted a new experimental trial which compares our approach with the perspective-based approach as reflected by Carver et al.'s contribution~\cite{carver2006finding}. The results of this new study strengthen our confidence in improving effectiveness and efficiency when using our approach. 

The remainder of this work is organized as follows. Section~\ref{sec:background_relatedWork} introduces the background of agile RE and how security verification is typically performed in this context. In this section, we also present related work. Section~\ref{subsec:carver_approach} presents in detail the technique we chose to evaluate our proposed approach, in this case, the perspective-based reading proposed by Carver et al.~\cite{carver2006finding}. Section~\ref{sec:our_approach} introduces the approach we designed to deal with security verification in agile contexts. Section~\ref{sec:experiment} presents the study design used to evaluate the approach. The results of the experimental trials are presented in Sections~\ref{sec:results}. We discussed threats to validity and the results of the experiments in Section~\ref{sec:threatsValidity} and~\ref{sec:discussion}, respectively. Section~\ref{sec:limitations} provides a discussion of limitations of our approach. Finally, our concluding remarks are presented in Section~\ref{sec:concluding_remarks}.

\section{Background and Related Work} \label{sec:background_relatedWork}

This section introduces the background on agile RE, inspections based on reading techniques, security properties and requirements and the synergy between these fields with verification activities. In addition, we describe related work to our proposed approach.

\subsection{Agile Requirements} \label{subsec:agile_requirements}

The term \say{agile requirements} emerged in response to the agile manifesto. It is used to define the \say{agile way} of executing and reasoning about RE activities~\cite{inayat2015systematic}. Yet, not much is known about the challenges posed by the collaboration oriented agile way of dealing with RE activities. Ramesh et al.~\cite{ramesh2010agile} performed a study with 16 organizations that develop software using agile methods. They identified that agile RE practices resulted in challenges regarding neglected NFRs, minimum documentation and no requirements verifications. The recent report from the \emph{Naming the Pain in Requirements Engineering} initiative~\cite{fernandez2015naming} (NaPiRE) extends already known challenges with (i) communication flaws between teams and customers, and (ii) under-specified requirements that remain too abstract and, thus, are not measurable. These observations give a picture on the difficulties of dealing with NFRs in agile contexts. It is reasonable to believe that security requirements are no different in this respect.

\subsection{Inspection Based on Reading Techniques}
\label{subsec:requirements_verification}

Software inspection is a quality assurance method to detect defects early during the software development process. The aim is to guarantee that developers deal with complete, consistent, unambiguous, and correct artifacts. In general, several authors have worked on quality assurance methods for verifying the quality properties of requirements. One of the most compared and evaluated methods, in several experiments and studies, are defect detection techniques, so-called reading techniques~\cite{halling2001using}. Software reading techniques attempt to increase the effectiveness of individual reviewers by providing guidelines that can be used to examine (by reading) a given software artifact and identify defects~\cite{travassos1999detecting}. There is empirical evidence that software reading is a promising technique for increasing software quality for different situations and documents types~\cite{shull1998developing}.

Perspective-based inspection is a variant of a formal technical review. This type of inspection is based on explicitly defining the important stakeholders for a particular artifact and the types of issues that are of importance to the team. Rather than asking each reviewer to search for all types of problems, the perspective-based approach requires inspectors to examine the document using a role-based scenario based on how one specific stakeholder~\cite{carver2006finding}. For example, an inspection of system requirements may include an inspector using a tester perspective. This inspector reviews the requirements by following a scenario in which he/she considers how to generate test cases based on the requirements.

Weak alignment of RE with inspection activities may lead to problems in delivering the required products in time with the right quality~\cite{bjarnason2014challenges}; for instance, weak communication of requirements changes to testers may result in lack of verification of new requirements and incorrect verification of old invalid requirements, leading to software quality problems, wasted effort and delays~\cite{kraut1995coordination}.

\subsection{Security Properties and Requirements} \label{subsec:security_requirements}

Security is an important quality characteristic of any software system that manages valuable data and processes~\cite{azuma2001square}~\cite{chung2012non}~\cite{devanbu2000software}. Hence, software development should be conducted with security in mind at all stages and it should not be an afterthought~\cite{mcgraw2006software}. However, developing secure software is not a trivial task often due to lack of awareness and security expertise in stakeholders and the inadequacy of methodologies to support developers who are not security experts~\cite{houmb2010eliciting}. Typically, security is often dealt with in retrospective and retrofitted when the system has already been designed and put into operation~\cite{mellado2010systematic}. This causes defects that have a major impact on the project resulting in a higher cost to fix them.

Security properties are the targets the customer establishes for their security program. Without security properties, they do not know what they are trying to accomplish for security and therefore will not reach any goals~\cite{peine2008security}.
Security Requirements (SRs) engineering can provide a foundation for developing secure systems. Nevertheless, like other quality requirements, they tend not to have simple yes/no satisfaction criteria. Haley et al.~\cite{haley2008security} present some challenges related to SRs. First, people generally think about and express SRs in terms of \say{bad things} (negative properties) to be prevented. It is difficult, if not impossible, to measure negative properties. Second, for SRs, the tolerance on \say{satisfied enough} is small, often zero, given the implications of noncompliance. Moreover, stakeholders tend to want SRs satisfaction to be very close to yes. Third, the effort stakeholders might be willing to dedicate to satisfying SRs also depends on the likelihood and impact of a failure to comply with them. In recent years, SRs has been investigated by several researchers. Mellado et al.~\cite{mellado2010systematic} have conducted a systematic review of SRs approaches to summarize existing methodologies. Fabian et al.~\cite{fabian2010comparison} also provide a comparison of SRs methods. Methods such as SQUARE~\cite{mead2005security} and Microsoft SDL~\cite{howard2006security} are compared. 

\subsection{Related Work}
\label{subsec:security_verification}

Our related work was based on findings from literature reviews such as~\cite{alsaqaf2017quality, villamizar2018systematic} and empirical searches in indexed databases. For instance, Alsaqaf et al.~\cite{alsaqaf2017quality} shows that very little is known about the evolution of NFRs (included security) in agile software development setting and more research is needed to understand the contexts in which these approaches would fit and add value, specially because very little empirical evaluation has been conducted. These studies revealed an important number of published proposals addressing security requirements in agile context. Most of them focused on analysing, identifying and prioritizing these kind of requirements. On the other hand, we also found studies focused on showing challenges and practitioners perspectives in this context. However, few authors have addressed the specific problems of security verification activities (~\cite{carver2006finding}, ~\cite{elberzhager2009software},~\cite{peine2008security}). The picture is even poorer in the agile context as concluded in~\cite{villamizar2018systematic}. We are aware of only one study that explicitly states to address security verification activities in agile methods~\cite{domah2015nerv}. As far as security verifications are concerned, we include, as related work, some studies that do not explicitly mention their suitability in agile contexts, but given their domain application we consider them suitable to address the same direction of our approach.

Domah et al.~\cite{domah2015nerv} propose a lightweight methodology to address NFRs early in agile software development processes. NFRs elicitation, reasoning, and validation are considered within that methodology. Regarding verification, it depends on a quantification taxonomy with different levels of decomposition for identifying quantified validation criteria for each NFR. However, this methodology does not offer specific guidance to support inspectors in identifying security-related defects in requirements specifications. Hence, previous knowledge on security is required to take advantage of the methodology.

Riaz et al.~\cite{riaz2014hidden} describes a tool-assisted process for identifying key attributes of sentences to be used in security-related analysis and specification of functional security requirements using a set of context-specific templates. The tool takes natural language requirements artifacts (requirement specifications, feature requests, etc.) and a trained classifier for the current problem domain as input. It also parses the artifacts as text sentences and identifies which (if any) security properties relate to each sentence. The tool then presents the user with a list of applicable security requirements templates for the identified properties. We can say this work conceives security in the same way as our approach. The difference is that they propose an approach for identifying security requirements whereas our approach focuses on detecting defects from security requirements already specified.  

Elberzhager et al.~\cite{elberzhager2009software} propose a model for security goals that involves guided checklists to support inspectors when checking security. They describe a step-by-step guide that results in questions to be checked by an inspector. This model is similar to our proposal because it works using a reading technique that supports the inspector on how to review security. However, there are differences. First, our approach focuses on verifying SRs in early stages, i.e., right after requirements specification and within agile requirement artifacts. Second, our approach addresses high-level SRs as defined by the Open Web Application Security Project (OWASP)~\cite{owasp}, which provides a well-known industry standard on security. Furthermore, our proposal involves classifying the defects found by inspectors, providing a better understanding of the distribution of the problems.

Peine et al.~\cite{peine2008security} propose a model named Security Goal Indicator Tree (SGIT) that maps negative and non-local goals to positive, concrete features of the software that can be checked during an inspection. The model supports inspection of software documents from various phases of the development process. An SGIT links a security goal with numerous indicators (which may be beneficial or detrimental for the achievement of the goal) and structures the set of indicators by Boolean and conditional relationships enabling an efficient selection of indicator subsets. Despite the deep analysis provided by this work, it is not clear the level of expertise needed by inspectors to use the model. Furthermore, as the related work above, the model does not shed light on the type of defects detected.    

Carver et al.~\cite{carver2006finding} focus their proposal on a PBR technique with the aim of identifying security defects. They describe a set of perspectives that provide security-specific questions for a requirements inspection. Two of them are part of the PBR technique (designer and tester). They also created a new perspective based on the needs of a black hat tester. In this additional perspective, the reviewer focuses on three types of security information: cryptography, authentication, and data validation. According to the authors, those types of information and the related questions were adapted for requirements from Araujo and Curphey’s article on security code reviews~\cite{Araujo2005}. However, due to the large number of software vulnerabilities and the variety of ways to deploy computer attacks, it could not be enough to consider only three types of security controls. Indeed, the list is incomplete when compared to other security standards such as OWASP~\cite{owasp} or the Common Criteria~\cite{mellado2007common}.     

To summarize, only few approaches exist that address the systematic detection of security defects, especially during early stages, and the number decreases when considering agile contexts which redraws the picture of how security could and should be dealt with~\cite{villamizar2018systematic}. However, we consider this last work, proposed by Carver et al.~\cite{carver2006finding}, the most related work given its nature to detect defects early via a reading technique and its focus on security. For that reason, we decided using this work to compare it against our approach with the aim of evaluating the suitability with respect to others proposals. In the following section, we present in detail the work proposed by Carver et al.

\section{Perspective-Based Reading Black Hat Approach}
\label{subsec:carver_approach}

One of the objectives of Carver et al.'s work is to integrate practices from the security engineering and software engineering communities into a set of techniques for identifying and removing security vulnerabilities early in the software lifecycle. This work is an adaptation of the PBR technique to address the security vulnerability problem during a requirements inspection process. The authors tailored PBR to focus on software security by augmenting two of the standard PBR perspectives (the designer and the tester) with additional security specific questions. In addition, the authors proposed a new perspective based on the needs of a black hat tester. Using this approach, the inspector reviews the requirements by following a scenario in which he/she considers how to generate test cases based on the requirements. 

\subsubsection{Set of Perspectives for Requirements Inspection}

The \textbf{designer} perspective has the goal of ensuring that there is enough, consistent information present in the requirements to successfully create a system design. The existing scenario is augmented with questions that focus on whether important security-related information has been correctly specified rather than being left up to the designer, who may not be familiar with all details of the security policy. This perspective provides a set of questions that the reviewer should consider when following this perspective:

\begin{itemize}
    \item Have the requirements specified enough information about the security policies for the designer to understand whether a layered security policy is required instead of a single point of vulnerability?
    
    \item If several administrator roles are defined, have they been defined as separate accounts with limited access to security resources, or a single account with comprehensive super user permissions?
\end{itemize}

On the other hand, the scenario for the \textbf{tester} perspective remains unchanged, but is augmented with security specific questions. The inspector using the tester perspective has the goal of ensuring that the trustworthiness of the system will be knowable during the testing phase. The questions of this perspective should consider the following:

\begin{itemize}
    \item Have the requirements specified appropriate exception-handling functionality?

    \item Have the requirements specified adequate safeguards that would take effect once a malicious user has gained unauthorized access to the system?
    
    \item Does the system have a well-defined status, either a secure failure state or the start of a plausible recovery procedure, after a failure condition?
\end{itemize}

As a new perspective proposed by the authors, the \textbf{black hat} perspective focuses the reviewer on finding weaknesses in the requirements that could be exploited via an attack. The scenario that the reviewer follows is to create a set of malicious attack scenarios that seek to exploit system vulnerabilities. 

\subsubsection{Types of Security Properties}

While creating the black hat scenario, the reviewer focuses on three types of security properties at the requirements stage: Cryptography, Authentication/Authorization, and Data Validation. These types of properties, along with the related questions were adapted for requirements from Araujo and Curphey’s article on Security Code Reviews~\cite{Araujo2005}.

\textbf{Cryptography} relates to the encoding mechanisms specified for data items within the system. During the review, the inspector is looking for under-specified or incorrectly specified features that could be exploited. The questions include the following:

\begin{itemize}
    \item Can the encoding mechanism specified for transmission and storage of data be broken?
    
    \item Do the cryptography mechanism specified follow well-known, well-document- ed, and publicly scrutinized algorithms, and if not, can they be easily broken?
\end{itemize}

\textbf{Authentication/Authorization} focuses the reviewer on on determining how unauthorized users could gain access to the system. The questions include the following:

\begin{itemize}
    \item Can the protocols for validating user identity be broken?
    
    \item If account lockout is specified, are there requirements in place to prevent denial-of-service attacks?
    
    \item Can user privileges be artificially elevated due to omission or poorly specified requirements?
\end{itemize}

Lastly, \textbf{Data Validation} is an important source of security vulnerabilities and focuses the reviewer on determining whether invalid data could be entered into the system. The question is: Do the requirements leave any opportunities for invalid data to be entered by the lack of validation of external data?

\section{Our Approach} 
\label{sec:our_approach}

In this section, we present our approach for reviewing security-related aspects in agile requirements specifications of web applications. The approach was designed considering user stories and their security specifications as input and involves applying Natural Language Processing (NLP). The goal is to relate those user stories to candidate security properties and high-level security requirements proposed by the Open Web Application Security Project (OWASP), a well known online community that produces freely-available articles, methodologies, documentation, tools, and technologies in the field of web application security~\cite{owasp}. As a result, the approach provides a user story focused reading technique that can be used to support the manual inspection of agile requirements. The reading technique helps to verify the user story security specifications against the OWASP high-level security requirements to identify defects such as omissions, ambiguities, inconsistencies, or incorrect facts. In the following, we provide more details of the conception of the approach. 

\subsection{Assumptions} \label{subsection:assumptions}

The approach was designed with some underlying assumptions in mind. These assumptions are as follows.

\textit{Requirements are specified in a user story format}. The software industry has gradually increased the use of agile and hybrid methods in its projects~\cite{kuhrmann2017hybrid}. In this context, user story is the most frequently used artifact for requirements specification~\cite{schon2017agile}. Therefore, the approach is focused on agile and, more specifically, on user stories. The stories are often analyzed independently and structured in a sentence as follows: As a [\textit{role}], I want to [\textit{feature}], so that [\textit{reason}].

\textit{The OWASP represents a reliable baseline and standard of security guidelines}. OWASP has a strong focus on web applications, one of the targets of our approach. OWASP concerns providing practical information about security in web applications to individuals, corporations, universities, government agencies, and other organizations worldwide. Many open source security-related tools (e.g., SonarQube~\footnote{https://docs.sonarqube.org/latest/user-guide/security-rules/}) and current research (e.g.,~\cite{subashini2011survey}) on web application security use OWASP as a definitive reference. Hence, we consider the reliability of this project as a reasonable assumption. 

\subsection{Approach Scope Delimitation} \label{subsection:delimitation_approach}

Hereafter, we answer some potential questions to provide further understanding of the intended approach.

\textit{To whom is the approach intended?} Our approach was designed to support novice inspectors and junior security analysts. This work provides them with a reading technique to assist in the identification of defects related to security aspects in agile requirements specifications. According to Nerur~\cite{nerur2005challenges}, people with a high-level of competence are of vital importance in agile teams because they tend not to depend on documentation to fulfill their functions. Much of the knowledge in agile development is tacit and resides in the heads of the development team members~\cite{boehm2002get}. Even more challenging, competence in software security is typically not widely spread among agile practitioners~\cite{goertzel2007software}. Given this, it is helpful to guide novice inspectors, including junior security analysts, by providing a detailed reading technique to support them. We believe that senior security analysts could still use the approach, but they are outside of the scope of our evaluation.

\textit{What security-related aspects does the approach cover?} We decided to focus on security properties and high-level SRs as proposed by the OWASP~\cite{owasp}. These high-level SRs describe the most important security features that architects and developers should include in every web application~\cite{owasp}. The System and Software Quality Requirements and Evaluation (SQuaRE) model~\cite{zubrow2004software} also define security characteristics, which hereafter, for term compatibility, will also be referred to as security properties. OWASP contains three security properties: confidentiality, integrity, and availability. SQuaRE, in contrast, contains five: confidentiality, accountability, integrity, non-repudiation, and authentication. Based on their definitions, all of the SQuaRE security properties can be mapped onto the OWASP security properties. For our final list of considered security properties, we used the OWASP properties with a single change, splitting confidentiality into two separate properties: (i) confidentiality and (ii) identification and authentication. Table~\ref{tab:security_properties} presents the security properties considered by our approach. 

\begin{table}[H]
\caption{Security properties considered by our approach}
\label{tab:security_properties}
\begin{tabular}{l p{8cm}}
\hline\noalign{\smallskip}
Security Property  & \multicolumn{1}{l}{Description}\\
\noalign{\smallskip}\hline\noalign{\smallskip}
\multirow{1}{*}{Confidentiality (C)}     &  Degree to which the data is disclosed only as intended.\\
\multirow{1}{*}{Integrity (I)}  &  Degree to which a system or component prevents unauthorized access to, or modification of, computer programs or data.\\
\multirow{1}{*}{Availability (A)}   &  Degree to which a system or component is operational when required for use.\\
\multirow{1}{*}{\begin{tabular}[l]{@{}l@{}}Identification\\Authorization (IA)\end{tabular}} &  Degree to which the identity of a subject or resource can be proved to be the one claimed. 
\\\noalign{\smallskip}\hline
\end{tabular}
\end{table}

\textit{What types of requirements defects does the approach cover?} In RE, a defect can be defined as any problem of correctness and completeness with respect to the requirements, internal consistency, or other quality attributes~\cite{travassos1999detecting}. A common defect taxonomy used when inspecting requirements is the one proposed by Shull~\cite{shull1998developing}. The defect types in this taxonomy are: omission, ambiguity, inconsistent information, incorrect fact, and extraneous information. However, we excluded the extraneous information defect type (which concerns specifying requirements that are not needed). This decision was taken because we use the OWASP high-level SRs as a reference; while they are stated as mandatory for inclusion, they are not necessarily complete, given that specific security needs may sprout for specific applications. Hence, given the impact that a missing security requirement can have on the application, we did not feel comfortable to recommend exclusions. Table~\ref{tab:Defect_Type} shows the defect types covered by our approach, their definitions and examples applied to security.

\begin{table}[H]
\caption{Defect types definition and examples in scope of our approach}
\label{tab:Defect_Type}
\begin{tabular}{l p{4cm} p{4cm}}
\hline\noalign{\smallskip}
Defect Type          & \multicolumn{1}{l}{Definition} & Applied to Security\\
\noalign{\smallskip}\hline\noalign{\smallskip}
\multirow{1}{*}{Omission (O)}      & Necessary information about the system has been omitted from the software artifact. 
& One or more security requirements that are not covered by the specifications originally created by an agile team.           
\\
\multirow{1}{*}{Ambiguity (A)}    & A requirement has multiple interpretations due to multiple terms for the same characteristic.  & For example, \say{the system shall inactivate a session when it exceeds certain periods of inactivity} is ambiguous because it is not clear the amount of time necessary to inactivate the session. It could be seconds, minutes or hours.       
\\
Inconsistency (IS) & Two or more requirements are in conflict.      & One security requirement specifies to encrypt data with RSA algorithm but another one specifies to encrypt it with AES.                                
\\ 
\multirow{1}{*}{Incorrect Fact (IF)}          & A requirement asserts a fact that cannot be true under the conditions specified for the system. 
& For example, \say{the system shall protect the firewall} does not make sense because it is the firewall that protects the system.  
\\\noalign{\smallskip}\hline
\end{tabular}
\end{table}

\textit{What kind of review technique does the approach use?} Typically, developers and software analysts rely on ad-hoc methods or checklists to analyze documents. In an ad-hoc review, the reader is not given directions on how to read. The result is that reviewers tend to build up skills in document understanding slowly based on individual experiences acquired over time~\cite{basili1996studies}. For this reason, we decided to focus the review of our approach on a reading technique to increase the effectiveness of individual reviewers by providing a systematic guide that can be used to examine, in our case, security-related aspects and consequently to identify defects.

\textit{To which part of the lifecycle of agile methods can the approach be applied?} Agile methods are characterized by having iterative structures that should allow early delivery, continual improvement, and rapid and flexible response to change~\cite{beck2001manifesto}. Hence, we envision that our approach is used just before a user story is defined as ready for codifying. In this way, we avoid rework that can be caused by not considering a security control or integrating one requirement with another one.

\subsection{Overview of our Approach} \label{subsection:overview_approach}

\begin{figure}[H]
    \includegraphics[width=1\textwidth]{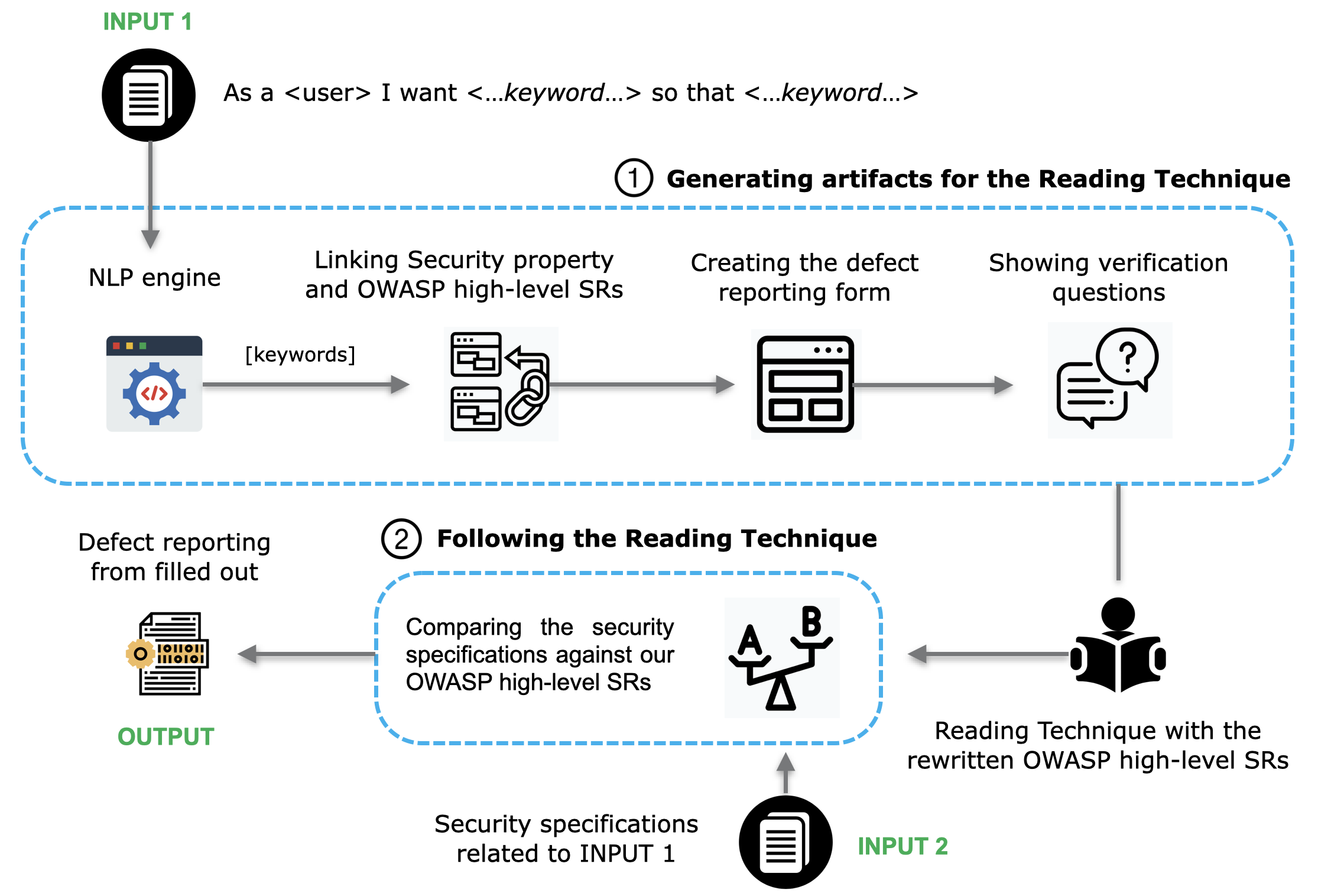}
    \caption{Overall structure of our approach}
    \label{fig:approach_overview}
\end{figure}

We propose our approach in two defined phases: (1) generating artifacts for the reading technique based on the agile requirements specification, and (2) following the reading technique to identify defects. Figure~\ref{fig:approach_overview} shows the flow and relationships between the artifacts and phases that form our approach. 

\subsubsection{Phase 1: Generating Artifacts for the Reading Technique} 

To generate the artifacts to follow our reading technique, we use Natural Language Processing (NLP) to extract keywords from the user story. Thereafter, these words are used to identify security properties and to link the related OWASP high-level SRs to be verified. The availability of automatic tools for the quality analysis of natural language requirements is recognized as a key factor for achieving
software quality~\cite{lami2004automatic}. Details on how keywords and security properties are identified follow. 

\textbf{Extracting keywords}. This activity involves automatically analyzing a user story that describes the features and functional requirements of the software to be built. Our approach extracts the relevant verb (action) of the user story that indicates a potential behavior to consider when thinking about security. For instance, the verb \say{export} could indicate confidentiality concerns because it is an action about transporting something. On the other hand, the verb \say{modify} could indicate integrity concerns because it is an action that involves altering something. In some cases, the nouns of the user story can also indicate situations where certain security features should be considered. This is particularly important to identify availability needs, e.g., time values (day, hour, second, period) may indicate scale/performance restrictions of the software. In that sense, nouns are also extracted for matching purposes. In summary, the verb is extracted from the second block of the user story format and nouns are extracted from the third block. Table~\ref{tab:words_extracted} shows where the words come from.

\begin{table}[H]
\caption{Way to extract the keywords from the user story}
\label{tab:words_extracted}
\begin{tabular}{p{2cm} p{5cm}}
\hline\noalign{\smallskip}
Type of Word  & \multicolumn{1}{l}{User Story Skeleton}  
\\\noalign{\smallskip}\hline\noalign{\smallskip}
Verbs   &  As a [user], I \textbf{[want to]}, [so that].
\\
Nouns   &  As a [user], I [want to], \textbf{[so that]}.
\\\noalign{\smallskip}\hline
\end{tabular}
\end{table}

\begin{figure}[H]
    \includegraphics[width=0.8\textwidth]{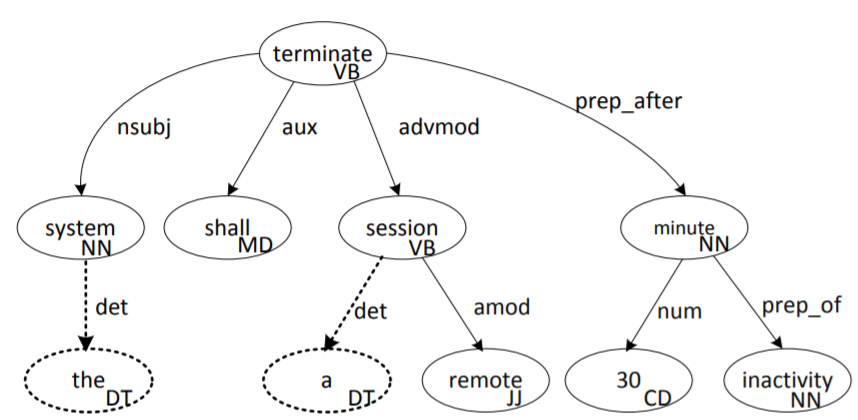}
    \caption{Sentence representation~\cite{slankas2013automated}}
    \label{fig:graph_nlp}
\end{figure}

To extract the words, we developed a Software Framework (FESRAS),\footnote{https://github.com/hrguarinv/FESRAS} which uses the Stanford CoreNLP tool\footnote{https://github.com/stanfordnlp/CoreNLP} through a library that provides a set of natural language analysis tools written in Java. The library represents each sentence as a directed graph where the vertices are words and the edges are the relationships between them. Thereby, the software framework can take the verbs and nouns of the user story. Figure~\ref{fig:graph_nlp} shows how the Stanford CoreNLP tool  represents the user story to identify verbs, nouns, among other kind of words. For this, consider the sentence: The system shall terminate a remote session after 30 minutes of inactivity. 

Note that the words extracted by using the Stanford Core NLP are not always keywords. After extracting the verb and nouns, we need to determine if these words match any keyword of our repository. If so, the user story contains at least one keyword that indicates some security concern that should be addressed. This matching is explained below. 

\textbf{Identifying Security Properties and Linking High-Level SRs}. After identifying the keywords of the user story, we need to identify security properties in order to map high-level SRs that represent a set of security-specific features to be verified. We use the keywords extracted from the user story to map security properties. Our software framework contains a set of keywords that are related to at least one security. Table~\ref{tab:keywords_secproperties} shows some these keywords that are part of our repository, to indicate which security properties should be considered. As an example, observe that the keyword \say{access}, that is a verb, indicates security concerns related to confidentiality, Identification and Authorization, because when accessing data, controls must be in place to guarantee the correct disclosure to it. Our online material, available at Zenodo\footnote{https://doi.org/10.5281/zenodo.3966542}, contains all the keywords of the repository.

\begin{table}[htb]
\caption{Relationship between the keywords and security properties}
\label{tab:keywords_secproperties}
\begin{tabular}{l l l l l}
\hline\noalign{\smallskip}
Keyword & Confidentiality & Integrity    & Availability & IA 
\\\noalign{\smallskip}\hline\noalign{\smallskip} 
Access                                                  & \multicolumn{1}{c}{X}   &                        &                        & \multicolumn{1}{c}{X}                    \\
Alter & & \multicolumn{1}{c}{X} & & \\
Apply &     &    &      &  \multicolumn{1}{c}{X} \\
Auto-populate &     & \multicolumn{1}{c}{X}    &      &   \\
Change                                                  &                          & \multicolumn{1}{c}{X} &                        &                                           \\
Create &     &  \multicolumn{1}{c}{X}  &      &   \\
Define &     &   &      & \multicolumn{1}{c}{X}   \\

Delete &     &  \multicolumn{1}{c}{X}  &      &   \\

Display & \multicolumn{1}{c}{X}     &    &      &   \\

Establish &     &    &      & \multicolumn{1}{c}{X}  \\

Export &  \multicolumn{1}{c}{X}   &    &      &   \\

Generate &     &  \multicolumn{1}{c}{X}  &      &   \\

Modify &     &  \multicolumn{1}{c}{X}  &      &   \\

Read &  \multicolumn{1}{c}{X}   &   &      & \multicolumn{1}{c}{X}  \\

Recover &     &    &  \multicolumn{1}{c}{X}    &   \\
Backup &     &    &  \multicolumn{1}{c}{X}    &   \\

Day &     &    &   \multicolumn{1}{c}{X}   &   \\

Hour &     &    &  \multicolumn{1}{c}{X}    &   \\

Password &     &    &      & \multicolumn{1}{c}{X}  \\

Period  &     &   &  \multicolumn{1}{c}{X}     &   \\

Privilege &     &    &      & \multicolumn{1}{c}{X}  \\

Role &     &    &      &  \multicolumn{1}{c}{X} \\

Time &     &    & \multicolumn{1}{c}{X}     &   \\
\noalign{\smallskip}\hline
\end{tabular}
\end{table}

This repository is based on a similar one provided by Slankas and Williams~\cite{slankas2013automated} in their work about automated extraction of NFRs in available documentation. However, we complement it by (1) considering additional keywords stated by the OWASP and (2) including synonyms of the words from these sources. By doing this, we increase the coverage of our repository. If there is no match between the words extracted and the keywords of our repository that indicates security properties, our approach will link the user story with their security specifications to all the security properties stored in our repository. With this the inspection will not be as advantageous in terms of effort and time as we anticipated because the reviewer will have to evaluate each of the security properties. However, in this way we ensure that the user story is examined by the four main domains of security in web applications.

After identifying the security property, we need to link the high-level SRs, which according to OWASP, are basic to deal with security in web applications. Table~\ref{tab:requirements_properties} shows the OWASP high-level SRs by security property. 

\begin{table}[htb]
\centering
\caption{OWASP high-level security requirements by security property}
\label{tab:requirements_properties}
\begin{tabular}{l p{8cm}}
\hline\noalign{\smallskip}
Security Property & OWASP High-Level Security Requirements
\\\noalign{\smallskip}\hline\noalign{\smallskip} 
\multirow{1}{*}{Confidentiality}  
& C1. Data shall be protected from unauthorized observation and disclosure both in transit and when stored.
\\ &  C2. System sessions shall be unique to each individual and cannot be shared.
\\ &  C3. System sessions are invalidated when timed out during periods of inactivity.
\\ &  C4. TLS protocol shall be used where sensitive data is transmitted. 
\\ &  C5. System shall use strong encryption algorithm at all times.      
\\
\multirow{1}{*}{Integrity}    
& I1. Any unauthorized modification of data must yield an auditable security-related event. 
\\ &  I2. All input is validated to be correct and fit for the intended purpose.
\\ &   I3. Data from an external entity shall always be validated.  
\\
\multirow{1}{*}{Availability}
&  A1. The application server shall be suitably hardened from a default configuration. 
\\ & A2. HTTP responses contain a safe character set in the content type header.
\\ &  A3. Backups must be implemented and recovery strategies must be considered.
\\
\multirow{1}{*}{\begin{tabular}[l]{@{}l@{}}Identification \&\\Authorization\end{tabular}}
& IA1. Users are associated with a well-defined set of roles and privileges. 
\\ & IA2. The digital identity of the sender of a communication must be verified.
\\ & IA3. Only those authorized are able to authenticate and credentials are transported and stored in a secure manner.
\\ & IA4. Passwords treatment must include complex passphrases, options to recover and reset the password and default passwords not allowed.
\\\noalign{\smallskip}\hline
\end{tabular}
\end{table}

\begin{figure}[htb]
    \includegraphics[width=0.8\textwidth]{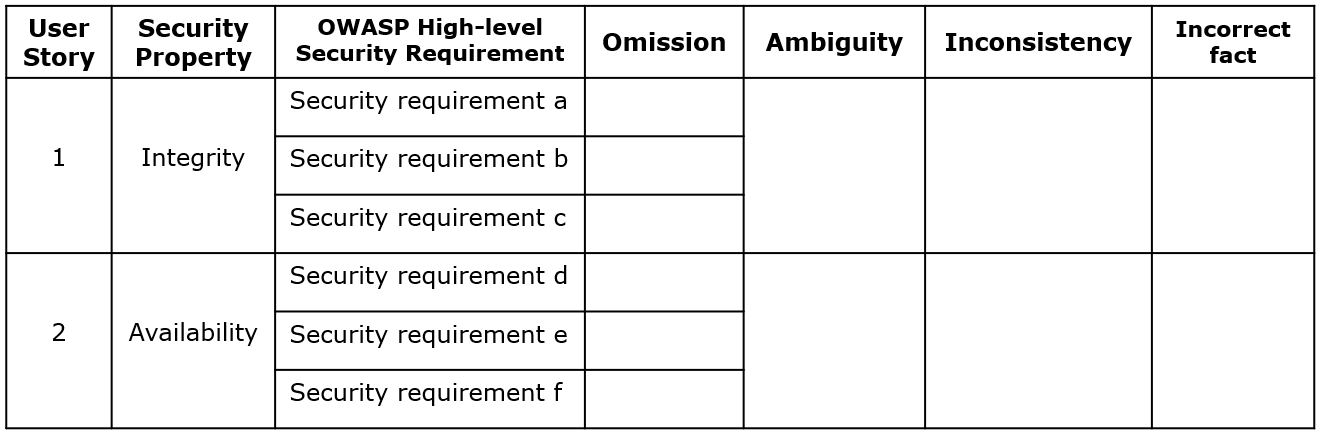}
    \caption{Defect reporting form}
    \label{fig:defect_reporting_form}
\end{figure}

At this point of our approach, we already identified the keywords of the user story, the security properties related to those keywords and consequently, the OWASP high-level SRs that address those security properties. The next step is to build the defect reporting form that works as a model and synthesizes most of the information the inspector needs to identify defects in the specifications. This form presents, in a structured way, much of the information that the reviewer must analyze to identify and classify the defects. Information such as the user story, the security property, the OWASP high-level SRs and the defect types are provided by the form. Figure~\ref{fig:defect_reporting_form} shows a template of this form with two of the security properties covered by our approach.

With the defect reporting form ready to be used the inspectors can use it as a model to verify whether the security specifications built, in a realistic example, by requirements or security analysts, contains any of the defect types covered in this work. This happens when inspectors compare the security specifications against the OWASP high-level SRs presented in the reporting form. To reach this, our approach formulates a set of verification questions to help identifying the different defect types in the security specifications. Table~\ref{tab:verification_questions} shows the questions.

\begin{table}[H]
\caption{Verification questions for the different defect types}
\label{tab:verification_questions}
\begin{tabular}{l p{8.5cm}}
\hline\noalign{\smallskip}
Type of Defect  & Question
\\\noalign{\smallskip}\hline\noalign{\smallskip} 
\multirow{1}{*}{Omission} &  When comparing the security specifications with the OWASP high-level SRs, are there high-level SRs or characteristics that were not specified?
\\
\multirow{1}{*}{Ambiguity} &  Does any security specification allow for multiple interpretations?
\\
\multirow{1}{*}{Inconsistency} &  Are there two or more security specifications in conflict?
\\
\multirow{1}{*}{Incorrect Fact} &  Is there any security specification stating information that is not true under the conditions specified?
\\\noalign{\smallskip}\hline
\end{tabular}
\end{table}

The first question aims to identify omission defects. For this type of defect, we provide to inspectors our rewritten OWASP high-level SRs that work as a model. With this, we seek to identify which SRs, stated by OWASP as basic and essential, are missing in the security requirements that were specified in agile software projects. Note that in the worst case, few or no security requirements are specified, as typically occurs in this type of projects. Thus, by doing this comparison we can offer relevant insights to inspectors to identify this type of defect. To identify the remaining defects, we use their definitions in a clear and short way. Our intention is to provide agile support to increase the efficiency of the inspection task. We believe that inspectors can directly associate the concept of the defect with the actions that allow them to identify such defects. For instance, to detect ambiguity defects our reading technique asks the inspector if any security specification allows for multiple interpretations. Regarding inconsistency defects, the inspectors must focus on figuring out whether two or more security specifications are in conflict. In this way, the meaning of the defect seeks to guide the inspector on the detection of ambiguity, inconsistency and incorrect fact defects. 

\subsubsection{Phase 2: Following the Reading Technique}

This phase aims to guide the reviewer in finding the requirements defects. Using the artifacts generated in Phase 1, we propose a reading technique that inspectors can follow to answer the verification questions in order to look for defects. As presented in Section~\ref{subsec:requirements_verification}, software reading techniques help a reviewer to \say{read} a requirement artifact for the purpose of finding relevant information that gives specific and practical guidance for identifying, in this case, security-related defects in agile requirements specifications.

To facilitate the review, our approach rewrites the OWASP high-level SRs in such a way that inspectors can easily identify certain security aspects. For instance, we use the \say{AND} logical connector in capital letters to get the attention of the reader and indicate that both aspects must be considered to satisfy the high-level SR, e.g., we have the following to the first confidentiality high-level SR: \textit{C1. Data shall be protected from unauthorized observation or disclosure both in transit \textbf{AND} when stored}. In this case, if the specifications were well specified, they must consider security aspects related to data protection both in transit and in storage. Otherwise, there is an omission defect.

\begin{figure}[H]
    \includegraphics[width=0.9\textwidth]{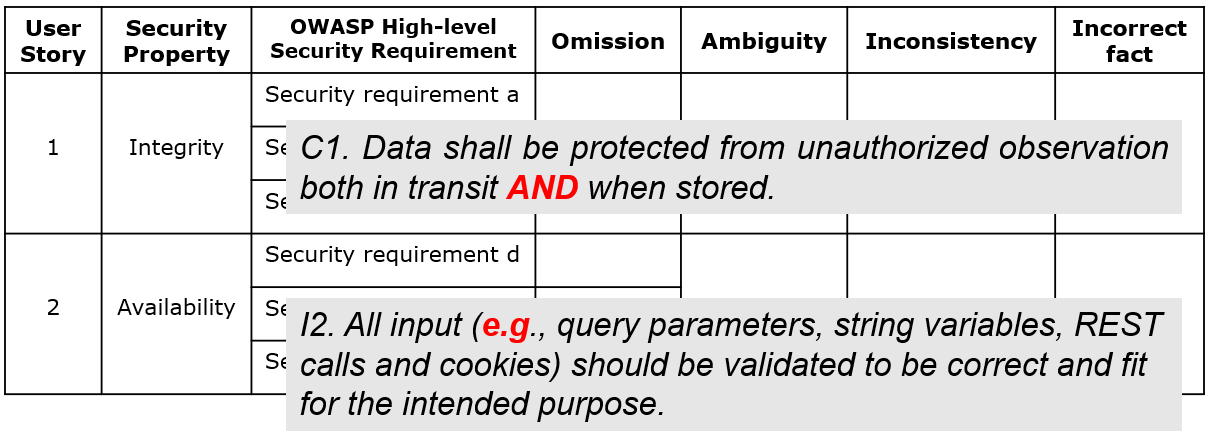}
    \caption{Visual aspect of the rewritten OWASP high-level SRs}
    \label{fig:rewriting_owasp}
\end{figure}

On the other hand, the security statements provided by our reading technique also present examples for some concepts in order to give inspectors an idea about the context of the OWASP high-level SRs. An example to the second integrity high-level SR follows: \textit{I2. All input, \textbf{e.g.}, query parameters, string variables and cookies, is validated to be correct and fit for the intended purpose}. In this way, reviewers are provided with a reading technique that should increase their performance during the review. Figure~\ref{fig:rewriting_owasp} shows an example of how the OWASP high-level SRs looks to get the attention of inspectors.

\begin{figure}[htb]
    \includegraphics[width=0.85\textwidth]{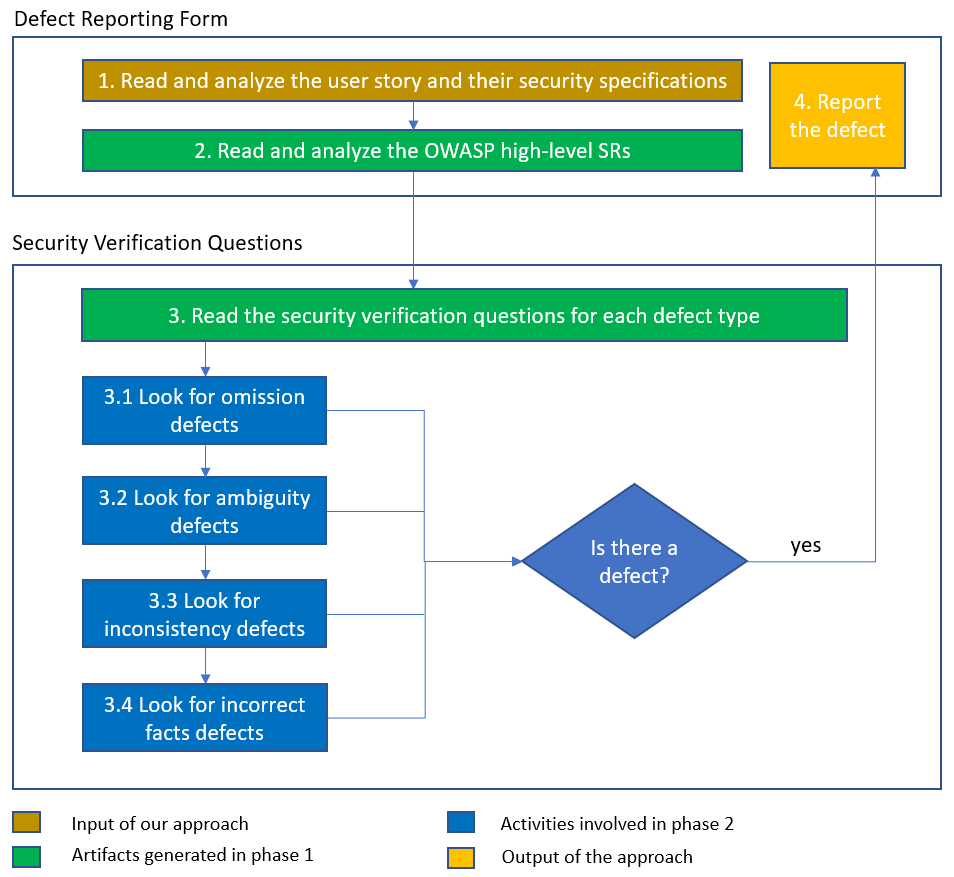}
    \caption{Procedures to apply to each user story}
    \label{fig:reading_approach}
\end{figure}

To summarize how inspectors should use our approach, we present Figure~\ref{fig:reading_approach} that provides a step-by-step of the activities to be followed by inspectors. The process begins by reading and analyzing the user story and their security specifications (input of our approach) to compare them with the OWASP high-level SRs. After this, inspectors can read our security verification questions for each defect type with the aim of detecting them and then report them (output of our approach). Please note that the process is repeated for each user story with their respective security specifications. We clarify that if there are no related security specifications as input in our approach, the result will be all of the OWASP high-level SRs related to the security property matched by the words extracted from the user story.

\subsection{Motivational Example} \label{subsection:motivational_example}

In the following, we demonstrate the application of our approach in an example setting. For this purpose, we present one of the agile specifications used in the experiments. Table~\ref{tab:user_example} shows a user story and its set of security specifications (inputs of our approach) with some defects commonly applied to a web application.

\begin{table}[H]
\caption{Input of the approach as agile requirements specification}
\label{tab:user_example}
\begin{tabular}{l p{7cm}}
\hline\noalign{\smallskip}
User Story & Security Specification (SS) 
\\\noalign{\smallskip}\hline\noalign{\smallskip} 
\multirow{10}{4cm}{As a customer, I want to be able to export my personal information so that I can use it in other systems.}  
& SS1. The system shall ensure that there is no residual data exposed.
\\
& SS2. The system shall store credentials securely using the AES encryption algorithm.
\\
& SS3. The system shall use the RSA encryption algorithm to protect all data all the time.
\\
& SS4. The system shall deactivate a session when it exceeds certain periods of inactivity. 
\\
& SS5. The system shall encrypt the roles and privileges of the system.
\\\noalign{\smallskip}\hline
\end{tabular}
\end{table}

With the user story in sight, the software framework extracts the words of the second and third block of the user story by using the Stanford CoreNLP, and then evaluates whether there is match to link the security properties. In this case, the extracted words are \say{export} (from the second block) and \say{system} (from third block). The following illustrates which keywords are extracted from the user story. 

\begin{quote}
    \textit{As a customer, I want to be able to \textbf{export} my personal information so that I can use it in other \textbf{systems}.}
\end{quote}

\begin{figure}[H]
    \includegraphics[width=0.8\textwidth]{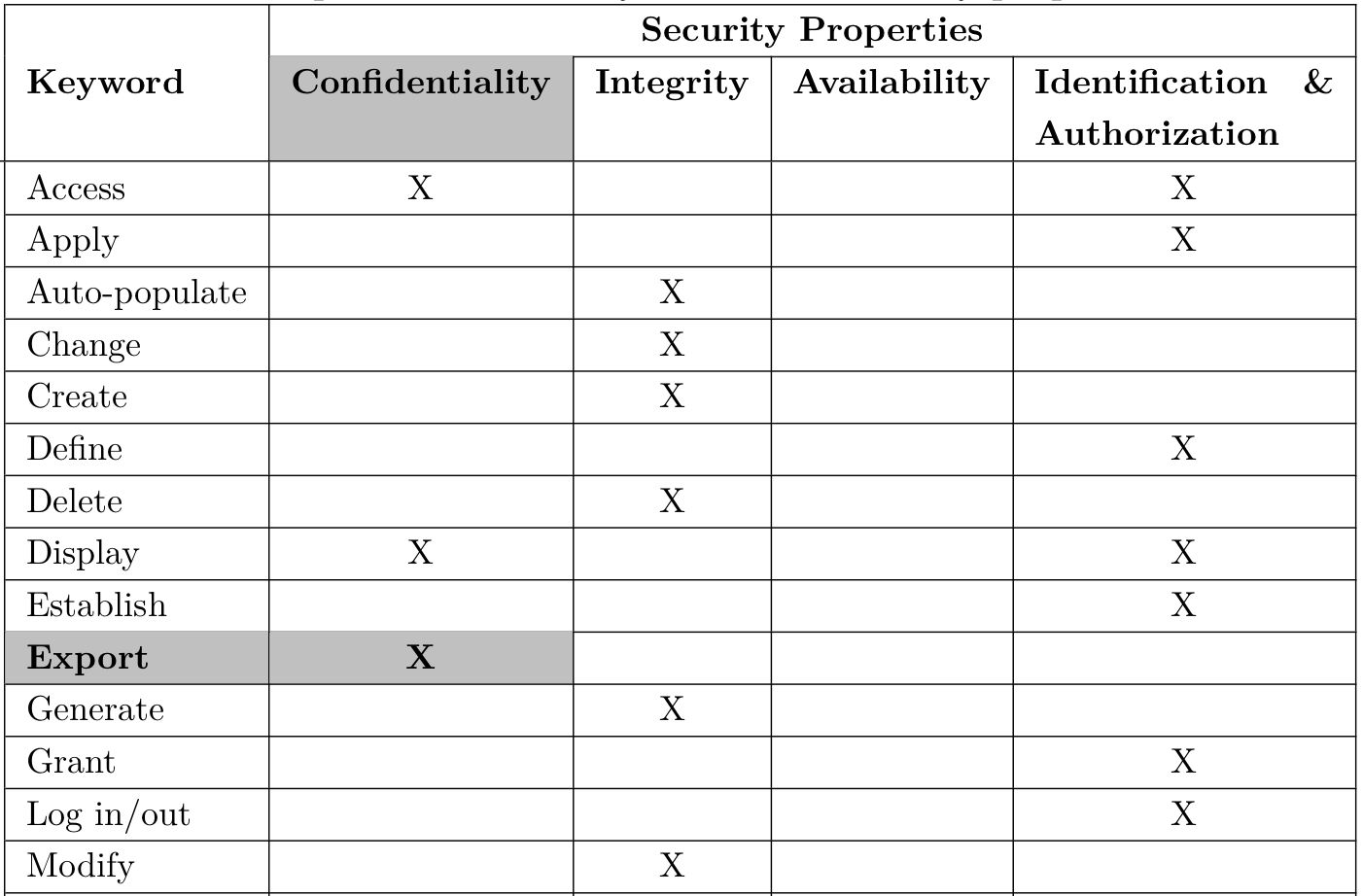}
    \caption{Matching keywords to security properties}
    \label{fig:linking_sr}
\end{figure}

Thereafter, the framework can verify whether some security property is related to the extracted words. According to Table~\ref{tab:keywords_secproperties}, \say{export} matches confidentiality, while \say{system} does not match any of the security properties. Figure~\ref{fig:linking_sr} presents the relationship between the keyword and the security property. 

In this way, since our approach identified confidentiality as security property, it can propose OWASP high-level SRs (\textit{Cf}. Table~\ref{tab:requirements_properties}, confidentiality). As part of our approach, we rewrite those OWASP high-level SRs to improve its readability and understanding. For that reason, the OWASP high-level SRs are provided to inspectors in the following way.

\begin{itemize}
  \item C1. Data shall be protected from unauthorized observation and disclosure both in transit \underline{AND} when stored.
  \item C2. System sessions shall be unique to each individual \underline{AND} cannot be shared.
  \item C3. System sessions are invalidated when timed out during periods of inactivity.
  \item C4. TLS protocol shall be used where sensitive data is transmitted. 
  \item C5. System shall use strong algorithms \underline{(e.g, DES, AES, RSA)} to encrypt data.
\end{itemize}

These OWASP high-level SRs are the basis to determine if the security specifications presented in Table~\ref{tab:user_example} contains omission defects. Note that inspectors are encouraged to read our verification questions to further analyze the quality of the specifications. 

Our approach then generates the defect reporting form by showing the user story with the linked security properties and OWASP high-level SRs. The verification questions are also provided here. Thus, inspectors know which security aspects they should verify. In that sense, the verification process starts at this point. By having inspectors responding to the verification questions looking for defects, we expect to obtain valuable insights from them on the quality of the security specifications. A sample enactment of answering these questions follows.

\textit{When comparing the security specifications with the OWASP high-level SRs, are there high-level SRs or characteristics that were not specified?} In this case, 3 out of 5 confidentiality high-level security requirements linked in Table~\ref{tab:requirements_properties} are related or make sense to the security specifications. However, we have two unspecified high-level requirements. Note that the second confidentiality high-level SR (C2) and the fourth confidentiality high-level SR (C4) are not covered by the security specifications. Therefore, we have detected two defects that should be marked as \say{omission}.

\textit{Does any security specification allow for multiple interpretations?} If we analyze with caution, we see that the fourth security specification (SS4) reflects a weak statement as the amount of time concerning \say{certain periods of inactivity}. It could be hours or seconds. Thus, we have identified a defect related to ambiguity.

\textit{Are there two or more security specifications in conflict?} The answer is affirmative. The second security specification (SS2) and third security specification (SS3) conflict because SS3 indicates to encrypt all data using the RSA algorithm. Nevertheless, SS2 indicates to protect credentials, which are also data, using the AES algorithm. Thus, we have identified a defect related to inconsistency.

\textit{Is there any security specification stating a characteristic that cannot be true under the conditions specified for the system?} The fifth security specification (SS5) is not correct because the concepts of the system cannot be encrypted. The action \say{encrypt} is not correct in the statement.

\begin{table}[H]
\caption{Defects reporting form}
\label{tab:defect_report}
\begin{tabular}{l|l|p{3.5cm}|l|l|l|l}
\hline
User Story & Security Property         &  OWASP High-Level SRs  & O & A   & IS      & IF \\\hline 
\multirow{17}{*}{US1}  & \multirow{18}{*}{Confidentiality} & C1. Data shall be protected from unauthorized observation AND disclosure both in transit AND when stored. &                   & \multirow{17}{*}{SS4} & \multirow{17}{*}{\begin{tabular}[c]{@{}l@{}}SS2\\SS3\end{tabular}} & \multirow{17}{*}{SS5} \\ \cline{3-4}
                    &                                  & C2. System sessions shall be unique to each individual AND cannot be shared.                              & \multirow{3}{*}{X}                 &                      &                            &                      \\ \cline{3-4}
                    &                                  & C3. System sessions are invalidated when timed out during periods of inactivity.                          &                   &                      &                            &                      \\ \cline{3-4}
                    &                                  & C4. TLS protocol shall be used where sensitive data is transmitted.                                       & \multirow{3}{*}{X}                 &                      &                            &                      \\ \cline{3-4}
                    &                                  & C5. System shall use strong algorithms (e.g, DES, AES, RSA) to encrypt data.                      &                   &                      &                            &                      \\ \hline
\end{tabular}
\end{table}

Finally, the reviewers fill out the defect reporting form that summarizes the defects found. Table~\ref{tab:defect_report} presents the output of the review using the reading technique. Note that the O column is related to the OWASP high-level SRs that were omitted to satisfy the Security Property (SP). The other columns are related to the remaining defect types.

In summary, this table indicates that the security specifications related to the user story contain six defects. Two out of them were marked as omission because the second and fourth OWASP high-level SRs related to confidentiality (C2, C4) are not covered by the security specifications. The rest of the defects (4) are related to ambiguous, inconsistent and incorrect fact defects. In this case, the fourth security specification (SS4) was marked as ambiguous, the second security specification (SS2) and the third security specification (SS3) were marked as inconsistent and the fifth security specification (SS5) was marked as incorrect.

\section{Experiment} \label{sec:experiment}

We evaluated the approach by conducting three controlled experiment trials with eight graduate and 48 undergraduate computer science students of the Pontifical Catholic University of Rio de Janeiro. The focus was to observe the impact of using our approach on effectiveness, efficiency, usefulness and ease of use.

We evaluated the study in two phases. In the first, we wanted to know whether our approach was suitable under certain non-complex conditions. With some positive results, the second phase was planned. In this case, the goal was to obtain more empirical evidence and then determine whether our approach is technically promising to use it in more complex environments. Thus, we conducted a new study by analyzing the effectiveness and efficiency of our approach when compared with the PBR Black Hat approach proposed by Carver et al.~\cite{carver2006finding}. This decision was supported due to that approach uses the same type of inspection technique that our approach and therefore is totally aimed at identifying security-related defects. 

The motivation for conducting the new study is to examine whether the findings can be replicated in different levels of support (e.g., using the PBR Black Hat approach), incorporating lessons learned from the first evaluation. We considered guidelines for reporting additional experimental studies proposed in~\cite{carver2010towards}. 

In this section, we detail the design of the experiments conducted to evaluate our approach. We present the goal, hypotheses, variable selection, selection of subjects, instrumentation, among others. We break down the experiments in three trials with different characteristics. In the original study, we allocated undergraduate and graduate students of Computer Science that were divided into two trial experiments. Regarding the new study (third trial), we assigned undergraduate students. Table~\ref{tab:experiment_trials} present some relevant details of the trials conducted across the study.

\begin{table}[H]
\caption{Controlled experiment trials conducted across the study}
\label{tab:experiment_trials}
\begin{tabular}{l l l l l l}
\hline\noalign{\smallskip}
Study & Trial & Date & Undergraduate  & Graduate & Total \\
\noalign{\smallskip}\hline\noalign{\smallskip}
Original& \multicolumn{1}{c}{1} & March, 2019& \multicolumn{1}{c}{25} & \multicolumn{1}{c}{0}  & \multicolumn{1}{c}{25}  \\
& \multicolumn{1}{c}{2} & March, 2019& \multicolumn{1}{c}{0} & \multicolumn{1}{c}{8} & \multicolumn{1}{c}{8}  \\
New & \multicolumn{1}{c}{3} & November, 2019 & \multicolumn{1}{c}{23} & \multicolumn{1}{c}{0} & \multicolumn{1}{c}{23} 
\\\noalign{\smallskip}\hline
\end{tabular}
\end{table}

In the following, we present similarities and differences between the original and the new study. In all the studies, subjects were assigned to the task of identifying security-related defects based on a set of agile requirements specifications. 

\subsection{Goal, Hypotheses and Research Questions} \label{subsec_experimentGoal}

For all the experiments (original and new study), we wanted to determine whether the use of our approach leads to efficient and effective detection of security-related defects when compared to an ad-hoc inspection based on personal expertise and the PBR Black Hat approach. Table~\ref{tab:GQM} details the definition of this study's goal by following the template provided by Basili~\cite{basili1992software}.

\begin{table}[H]
\caption{Goal definition of the experiments}
\label{tab:GQM}
\begin{tabular}{p{2.5cm} p{8.8cm}}
\hline
Analyze  &  the reading technique generated by our approach
\\
for the purpose of   & characterization
\\
\multirow{1}{*}{with respect to}   &  the effectiveness, efficiency, usefulness and ease of use of the approach
\\
from the point of view of   &  researchers (on the measured effectiveness and efficiency) and inspectors (on the perceived usefulness and ease of use)
\\
\multirow{1}{*}{in the context of}  &  novice inspectors using our approach, when compared to using the OWASP high-level SRs and the PBR Black Hat approach.
\\ \hline
\end{tabular}
\end{table}

Based on our goal, we formulated Research Questions (RQs) that seek to address two aspects that we believe should be covered by our approach. First, efficiency and effectiveness play an important role in inspection activities, as inspectors should be able to find defects (effectiveness) with reasonable effort (efficiency). Second, we also address usefulness concerns, because they are closely related to the adoption of the approach. Therefore, we defined the following two RQs that apply to both the studies conducted.

\begin{itemize}
    \item \textbf{(RQ1)} Does our approach have an effect on defect detection effectiveness and efficiency when compared to the other review approaches of the study?
    \item \textbf{(RQ2)} How do the inspectors perceive the usefulness and ease of use of our approach?
\end{itemize}

Based on RQ1, we derived hypotheses to be evaluated quantitatively. Note that we defined hypotheses that vary depending on the study performed and that they may refuted or supported only in comparison with the considered other methods. In the original study, the following hypotheses were derived:

\begin{itemize}
    \item \textbf{H$_{01a}$:} There is no difference in terms of effectiveness when using our approach, when compared to using the OWASP high-level SRs. 
    \item \textbf{H$_{11a}$:} There is a difference in terms of effectiveness when using our approach, when compared to using the OWASP high-level SRs.
    \item \textbf{H$_{02a}$:} There is no difference in terms of efficiency when using our approach, when compared to using the OWASP high-level SRs.
    \item \textbf{H$_{12a}$:} There is a difference in terms of efficiency when using our approach, when compared to using the OWASP high-level SRs.
\end{itemize}

Regarding the new study, we compare the performance of using our approach against the PBR Black Hat approach. Thus, we defined the following hypotheses.
\begin{itemize}
    \item \textbf{H$_{01b}$:}There is no difference in terms of effectiveness when using our approach, when compared to using the PBR Black Hat approach. 
    \item \textbf{H$_{11b}$:} There is a difference in terms of effectiveness when using our approach, when compared to using the PBR Black Hat approach.
    \item \textbf{H$_{02b}$:} There is no difference in terms of efficiency when using our approach, when compared to using the PBR Black Hat approach.
    \item \textbf{H$_{12b}$:} There is a difference in terms of efficiency when using our approach, when compared to using the PBR Black Hat approach.
\end{itemize}

\subsection{Variables Selection} \label{subsec_variablesSelection}

The independent variable is the treatment applied by the groups to detect defects in the SRs specifications. In that sense, inspectors who were part of the control group in the first and second trial received as support the OWASP high-level SRs. On the other hand, in the third trial, inspectors who were part of the control group received the PBR Black Hat approach, but at the same time, they received the same support that inspectors in the first and second trial, that is, they also received the OWASP high-level security requirements. With respect to the experimental group, all the inspectors received our approach. 

\begin{table}[H]
\caption{Metrics used to answer the RQs and test the hypotheses}
\label{tab:metrics}
\begin{tabular}{p{1.9cm} p{1.7cm} p{7cm}}
\hline\noalign{\smallskip}
Criteria & Type  & \multicolumn{1}{l}{Description}\\
\noalign{\smallskip}\hline\noalign{\smallskip}
Effectiveness &  Quantitative & Ratio between the number of defects found and the total of seeded defects in the specifications \\
Efficiency    &  Quantitative   &  Ratio between the number of real defects found and the time spent in finding them\\
Usefulness &  Quantitative & Frequency of the participants' perception on the usefulness of the approach using a follow-up questionnaire\\
           &  Qualitative & Coding of the answers of the follow-up questionnaire \\
Ease of use & Quantitative  & Frequency of the participants' perception on the ease of use of the approach using a follow-up questionnaire \\
           &  Qualitative & Coding of the answers of the follow-up questionnaire
\\\noalign{\smallskip}\hline
\end{tabular}
\end{table}

Regarding the dependent variables, we used effectiveness and efficiency, defined as follows. \textit{Effectiveness} is expressed as the ratio between the number of real defects found and the total of seeded defects in the documents. On the other hand, \textit{Efficiency} refers to the ratio between the number of real defects found and the time spent in finding them. For these variables, we collected quantitative data to test the hypotheses presented in Section~\ref{subsec_experimentGoal}. We also collected qualitative data with open questions via a follow-up questionnaire. The aim was to gain insights about the perceived usefulness and ease of use of the approach. Table~\ref{tab:metrics} summarizes the metrics collected in the experiments.

\subsection{Selection of Subjects} \label{subsec_selectionSubjects}

Our subjects were intended to represent novice inspectors. We thus selected subjects, by convenience, from classes on computer science at PUC-Rio, involving, for the original study, 25 undergraduate (first trial) and eight graduate students (second trial). Regarding the new study, the experiment was conducted in one trial (third trial), involving 23 undergraduate students from classes on computer science at PUC-Rio. There is evidence that using students is an effective way to advance software engineering theories and technologies but, like any other aspect of study settings, should be carefully considered during the design, execution, interpretation, and reporting of an experiment~\cite{falessi2018empirical}.

We characterized the subjects by their experience and knowledge in five areas: agile software development (ASD), agile RE (ARE), security aspects (SA), security requirements (SR) and requirements inspections (RI). To this end, we defined a scale from one to five where lower score indicates lower experience and high score indicates experience in the industry. 

Table~\ref{tab:characterization} shows details of the characterization of the students of the original study. Subjects with at least three values greater to three were highlighted to identify the participants were equally divided between the groups of the experiments, applying the blocking principle. We did not list subjects of the new study because none of them met the requirements to be treated differently. Therefore, they were randomly assigned into control and experimental groups. 

As a result, we found that the majority of students had a low level of security and requirements inspection experience. Hence, they match our intended profile (novice inspectors).

\subsection{Experiment Context} \label{subsec_experimentContext}

In the following, we detail the experiment context of the studies. Before conducting the controlled experiment of the first trial, we carried out a pilot study with two independent volunteers. The aim was to evaluate the overall (particularly technical) feasibility, time, adverse events, and improve the experiment materials before the experiment trials.

All the studies were conducted with the same agile specifications. These requirements contain a set of user stories in this format: As a [\textit{Role}], I want [\textit{Feature}], so that [\textit{Reason}]. The document also contained their related security specifications with seeded defects that represent specifications that in real settings, would be created by requirements analysts or product owners in agile teams. (cf. Section~\ref{subsection:motivational_example} to see one of the specifications used in the experiments).

\begin{table}[H]
\caption{Details of the characterization of the subjects}
\label{tab:characterization}
\begin{tabular}{l l c c c c c c c}
\hline\noalign{\smallskip}
Experiment & Level & ID & Trial & ASD  & ARE & SA & SR & RI \\
\noalign{\smallskip}\hline\noalign{\smallskip}
Original & Undergraduate & 1 & 1 & 3 & 3  & 1  & 1   & 2\\
        & & 2 & 1 & 2 & 1  & 1  & 1  & 2  \\
        & &\cellcolor[HTML]{C0C0C0}3 &\cellcolor[HTML]{C0C0C0}1 &\cellcolor[HTML]{C0C0C0}4 &\cellcolor[HTML]{C0C0C0}4 & \cellcolor[HTML]{C0C0C0}4  &\cellcolor[HTML]{C0C0C0}4  & \cellcolor[HTML]{C0C0C0}2 \\
        & &\cellcolor[HTML]{C0C0C0}4 &\cellcolor[HTML]{C0C0C0}1 & \cellcolor[HTML]{C0C0C0}4 &\cellcolor[HTML]{C0C0C0}1 & \cellcolor[HTML]{C0C0C0}4  &\cellcolor[HTML]{C0C0C0}3  &\cellcolor[HTML]{C0C0C0}4 \\
        & &\cellcolor[HTML]{C0C0C0}5 &\cellcolor[HTML]{C0C0C0}1 &\cellcolor[HTML]{C0C0C0}5 & \cellcolor[HTML]{C0C0C0}2 &\cellcolor[HTML]{C0C0C0}4  &\cellcolor[HTML]{C0C0C0}4  & \cellcolor[HTML]{C0C0C0}2 \\
        & & 6 & 1 & 2 & 2 & 2  & 1  & 2 \\
        & & 7 & 1 & 2 & 2 & 3  & 3  & 3\\
        & & 8 & 1 & 2 & 2 & 2  & 1  & 1\\
        & & 9 & 1 & 3 & 2 & 1  & 1  & 2\\
        & & 10 & 1 & 5  & 2  & 2   & 1   & 2 \\
        & & 11 & 1 &  2 & 1  & 2   & 2   & 1 \\
        & & 12 & 1 &  3 & 2  & 1   & 2   & 2 \\
        & & 13 & 1 &  1 & 1  & 1   & 1   & 1 \\
        & & 14 & 1 &  2 & 5  & 3   & 3   & 1 \\
        & & 15 & 1 &  2 & 2  & 2   & 2   & 2 \\
        & & 16 & 1 &  5 & 5  & 2   & 1   & 3 \\
        & & 17 & 1 &  3 & 1  & 1   & 1   & 1 \\
        & & 18 & 1 &  3 & 1  &  3  & 2   & 2 \\
        & & 19 & 1 & 4  & 1  & 1   & 2   & 2 \\
        & & 20 & 1 &  5 & 5  & 2   & 2   & 1 \\
        & & 21 & 1 & 4  & 4  & 1   & 1   & 1 \\
        & &\cellcolor[HTML]{C0C0C0}22 &\cellcolor[HTML]{C0C0C0}1 & \cellcolor[HTML]{C0C0C0}5  &\cellcolor[HTML]{C0C0C0}5  &\cellcolor[HTML]{C0C0C0}2  &\cellcolor[HTML]{C0C0C0}2  & \cellcolor[HTML]{C0C0C0}4 \\
        & & 23 & 1 & 2  & 3  & 2   & 2  & 2 \\
        & & 24 & 1 & 4  & 4  & 1   & 1   & 1 \\
        & & 25 & 1 & 2  & 2  & 3   & 2   & 1 \\
& Graduate & 26 & 2 & 3 & 2 & 3  & 3  & 1\\
        & & 27 & 2 & 2 & 2 & 3  & 3  & 1\\
        & &\cellcolor[HTML]{C0C0C0}28 &\cellcolor[HTML]{C0C0C0}2 &\cellcolor[HTML]{C0C0C0}4 &\cellcolor[HTML]{C0C0C0}2 & \cellcolor[HTML]{C0C0C0}4  &\cellcolor[HTML]{C0C0C0}4  &\cellcolor[HTML]{C0C0C0}2\\
        & & 29 & 2 & 4 & 2 & 1  & 1  & 1\\
        & & 30 & 2 & 4 & 1 & 2  & 2  & 2\\
        & &\cellcolor[HTML]{C0C0C0}31 &\cellcolor[HTML]{C0C0C0}2 &\cellcolor[HTML]{C0C0C0}4 &\cellcolor[HTML]{C0C0C0}3 & \cellcolor[HTML]{C0C0C0}4  &\cellcolor[HTML]{C0C0C0}4  &\cellcolor[HTML]{C0C0C0}5\\
        & & 32 & 2 & 2 & 2  & 2  & 3  & 3\\
        & & 33 & 2 & 2 & 2 & 2  & 2  & 2
\\\noalign{\smallskip}\hline
\\\multicolumn{5}{l}{%
}
\end{tabular}
\end{table}

Aiming at mitigating threats to validity concerning the distribution of subjects between groups, we characterized the subjects and then applied the principles of balancing, blocking and random assignment~\cite{wohlin2012experimentation}. In all the trials, students who demonstrated experience on the topics involved in the study were separated and distributed equally into the control and experimental group. In the case of the first and second trials, we allocated equally 6 out of 33 between the groups, given they had higher scores. Regarding the new study, subjects were randomly placed into the experimental and control group since the characterization did not shed light. Table~\ref{tab:agile_specifications} shows the support given to the control and experimental group in each trial.

In Table~\ref{tab:number_subjects_per_treatmeant}, we present the number of undergraduate students (US) and graduate students (GS) in each group of the trials. 

Subjects who found less than two defects were discarded as outliers because, in our understanding, their results reflect a lack of interest in having a good performance in the review. In the original study, we discarded two subjects from the first trial and one for the second trial. All these discarded subjects conducted the inspection by using the OWASP high-level SRs (control group). Regarding the new study, we discarded one subject from the third trial. In this case, the discarded subject conducted the inspection by using our approach (experimental group). 

\begin{table}[H]
\caption{Support provided to each group of the experiments}
\label{tab:agile_specifications}
\begin{tabular}{p{2.4cm} p{6cm} p{3cm}}
\hline\noalign{\smallskip}
Trial & Control Group  & \multicolumn{1}{l}{Experimental Group}\\
\noalign{\smallskip}\hline\noalign{\smallskip}
1, 2    & OWASP high-level SRs                    & Our approach \\
3 & PBR Black Hat + OWASP high-level SRs & Our approach
\\\noalign{\smallskip}\hline
\end{tabular}
\end{table}

\begin{table}[H]
\caption{Number of subjects by experiment, trial and group}
\label{tab:number_subjects_per_treatmeant}
\begin{tabular}{l l l l l}
\hline\noalign{\smallskip}
Experiment & Trial  & Control & Experimental & Total \\
\noalign{\smallskip}\hline\noalign{\smallskip}
Original    & 1 & 12 US  & 13 US & \multicolumn{1}{c}{25}\\
            & 2 & 4 GS   & 4GS  &  \multicolumn{1}{c}{8} \\
New & 3  & 11 US  & 12 US & \multicolumn{1}{c}{23}
\\\noalign{\smallskip}\hline
\end{tabular}
\end{table}

To avoid the defect seeding to represent a confounding factor, the type and amount of seeded defects to evaluate the suitability of our approach was carefully considered. All the trials were conducted with the same type and distribution of defects. Table~\ref{tab:seeded_defects} shows the distribution of the seeded defects per user story. In total, 14 defects were seeded. Three independent researchers reviewed the representativeness of the requirements specifications and the defects before conducting the experiment trials.

\begin{table} [H]
\caption{Distribution of the seeded defects}
\label{tab:seeded_defects}
\begin{tabular}{l l l l l l}
\hline\noalign{\smallskip}
User Story  & Omission & Ambiguity & Inconsistency & Incorrect Fact & Total    
\\\noalign{\smallskip}\hline\noalign{\smallskip}
US1   &  \multicolumn{1}{c}{2} & \multicolumn{1}{c}{2} & \multicolumn{1}{c}{2} & \multicolumn{1}{c}{1} & \multicolumn{1}{c}{7}
\\
US2   &  \multicolumn{1}{c}{2} & \multicolumn{1}{c}{2} & \multicolumn{1}{c}{2} & \multicolumn{1}{c}{1} & \multicolumn{1}{c}{7}
\\\noalign{\smallskip}\hline
\end{tabular}
\end{table}

“The original (first and second trial) and the new study (third trial) differed in terms of certain aspects related to the setting of the study. In the original study 33 subjects participated, divided into two trials with 25 undergraduate and eight graduate students. In the new study, we involved 23 undergraduate students. This number of participants was defined according to the availability of students to whom we had access and who met our study profile. We conducted the experiments as an in-class activity to have the attendance of most of our subjects. We motivated the subjects to give their best, while we guaranteed the confidentiality of their results. Regarding training, in the original study we did not provide training on inspection techniques because, at that time, we considered that our approach and our task description could be self-assimilated by the students. After the lessons learned from the first study, we identified that training should be provided in the new study. For that reason, we provided training on both, our approach and  the  PBR  Black  Hat  approach.  Additionally,  we  reminded  inspectors  to pay  attention  to  the  task  description  according  to  the  treatment  provided. These changes were motivated by the feedback received by the inspectors in the follow-up questionnaire of the first two trials. Table~\ref{tab:context_experiment} presents the main factors involved in the experiments to summarize the context of the original and the new study. 

\begin{table} [H]
\caption{Context factors of the experiments}
\label{tab:context_experiment}
\begin{tabular}{p{2.5cm} p{3.2cm} p{4.3cm}}
\hline\noalign{\smallskip}
Context Factors  & Original Study & New Study    
\\\noalign{\smallskip}\hline\noalign{\smallskip}
Subjects   & 25 US and 8 GS &  23 US \\\\
Setting   & In-class activity   &  In-class activity  \\\\
Training provided   &  --Security properties and OWASP \newline --Type of defects & --Security properties and OWASP \newline --Type of defects \newline --Our reading technique and PBR Black Hat approach \\\\ 
Other changes   & NA  & --Reminder to follow the task description \newline
\\\noalign{\smallskip}\hline
\end{tabular}
\end{table}

\subsection{Experiment Design} \label{subsec_experimentDesign}

\begin{figure}[htb]
    \includegraphics[width=0.9\textwidth]{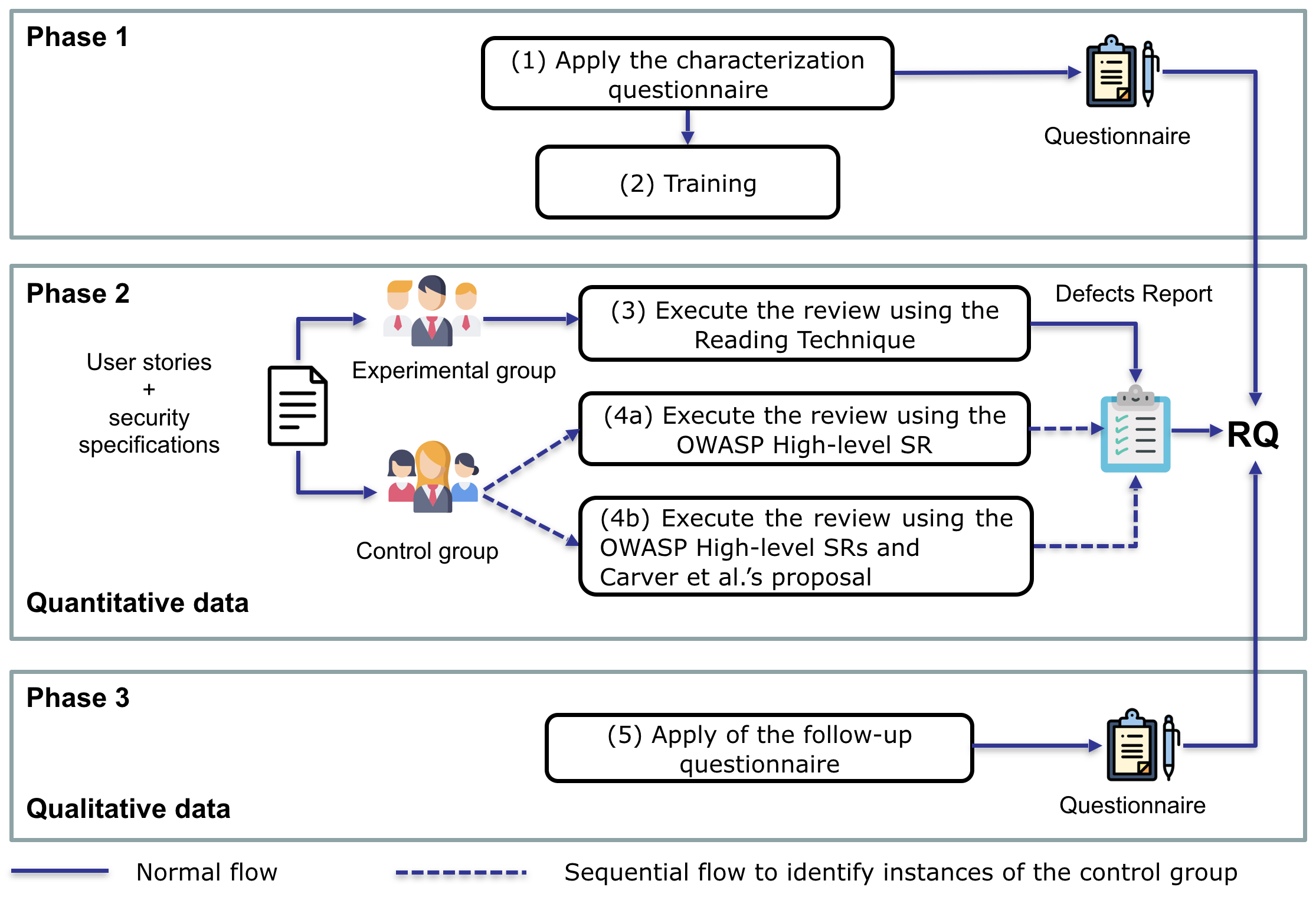}
    \caption{Experimental design of the study}
    \label{fig:experiment_desgn}
\end{figure}

Our experiments were composed of one factor with two treatments: (1) using our proposed reading technique and (2) using the OWASP high-level SRs (first two trials) or using the PBR Black Hat approach (third trial). Figure~\ref{fig:experiment_desgn} shows all the phases involved in the experiments.

The study design is composed of a set of artifacts distributed into three phases. Details of the artifacts used in the experiment are available at Zenodo\footnote{https://doi.org/10.5281/zenodo.3966542}. In the first phase, all the students filled out a characterization questionnaire with questions about their expertise in the topics related to the study. They also received training to introduce the main topics. In the second phase, we obtained quantitative data by conducting the original (first and second trial) and new study (third trial). The students of all the experiments were divided into two groups in order to evaluate the performance by executing the review using or not our approach. Finally, in the third phase, the participants of the experiments gave us feedback on the execution of the experiment. The instruments used to conduct the experiment
are further described in the following.

\subsection{Instrumentation} \label{subsec_instrumenation}

\textit{Characterization questionnaire}. The goal of the questionnaire is to characterize the experience of the subjects and identify key characteristics of four topics: agile software development, RE, software security and inspections.

\textit{Training}. In the first and second trials, the training was focused on the security properties, the OWASP high-level SRs and the defect types involved in the experiment. In the third trial, we included training on how to use each of the inspection techniques. Furthermore, we provided feedback on common errors along with the experiments and we emphasize strictly following the task description of the experiment.  

\textit{User story and security specifications:} The input of our approach is a user story and its set of security specifications. Therefore, we created the specifications following the guidelines proposed by Lucassen et al.~\cite{lucassen2015forging}. They were based on typical customer requests for developing web applications, e.g., sending sensitive information to other systems and deleting data. When doing so, we relied as an orientation on SRs specifications from real industrial software projects as used by our industry partners. We did not use real specifications verbatim in our setting for confidentiality reasons.

\textit{Task description}. This document explains to the students how to fill out the defect reporting form according to their treatment. Both treatments received the same specifications. For one treatment (experimental group), the technique was generated according to the user story and its related OWASP high-level SRs. For the other treatment (control group), in the first two trials, the list of the OWASP high-level SRs with the list of the defect types was provided. In the third trial the control group also received the PBR Black Hat approach.

\textit{Defect reporting form}. This form was used by subjects to record the start and end time of the review, as well as the defects by location and type. The defect reporting form for the experimental group was the one generated for applying the reading technique in Table~\ref{tab:defect_report}.

\textit{Follow-up questionnaire}. This questionnaire was based on the Technology Acceptance Model (TAM). TAM has been extensively used in several studies~\cite{turner2010does}. We wanted to know whether the approach was useful and easy to use. We included open text questions to gather feedback about the difficulties and benefits of using our approach.

\subsection{Data Collection Procedure} \label{subsec_data_collection}

We used the following procedures to collect the data necessary for answering our research questions and test our hypotheses.

\textit{Collect defects found and false positives:} Inspectors applied the reading technique generated by our approach on the given requirements artifacts, which generated a list of detected defects by each inspector. The same process was applied by the remaining treatments involved in the experiments. We reviewed the output of each inspector in pairs of researchers, and classified a registry as false positive if it was not a defect. Based on this data, we evaluated the performance of the treatments in terms of effectiveness, considering only defects (i.e., true positives).

\textit{Collect time spent for detecting defects:} In this study, time spent refers to the amount of time inspectors spent to identify defects. We collected the time spent by each inspector during the inspection. The time spent is the difference between the start time and end time. Note that this time does not include activities such as training and follow-up questionnaire. We defined 1 hour as the maximum time limit to be spent by an inspector to find the seeded defects. The pilot study was helpful to define the time it would take for inspectors to complete the inspection task. Based on this data, we evaluated the performance of the treatments in terms of efficiency.

\textit{Collect perceptions of the subjects: } Finally, to answer RQ2, we used the TAM based ~\cite{turner2010does} follow-up questionnaire to collect feedback on the usefulness and ease of use of our approach and the PBR Black Hat approach.

\subsection{Analysis Procedure}

We structure our analysis procedure into four steps. Each step leads to the results necessary for answering one of our research questions.

\textit{Calculate effectiveness and efficiency of the experiment treatments: } First, we analyzed the performance of each treatment by carrying out descriptive statistics such as the percent of defects detected in the agile specifications and the time duration to perform the review. Hypotheses testing was also applied. To this end, the analysis was conducted using the statistical tool RStudio version 1.1.4. For the hypotheses testing, we used the Mann-Whitney test with alfa = 0.05. The small number of independent samples motivated this choice of statistical significance and test. Second, to get a deeper insight into the defects detection, we analyze the distribution of types of defects identified by inspectors. This provided answers for RQ1.

\textit{Analyze the number of false positives: } We know that the reliability of an inspection technique is important. Therefore, we analyze false positives to determine to which extent our approach leads inspectors to false positives. We also analyzed this metric for the approaches compared in the experiments.

\textit{Interpret questionnaire answers: } We analyze the frequencies of responses to the TAM questionnaire. Additionally, we conducted a qualitative analysis applying grounded theory open coding activities to the open questions. This provided answers to RQ2.

\subsection{Operation of the Experiments} \label{subsec_experimentOperation}

The original study (i.e., the first two trials) was conducted in March of 2019 ~\cite{villamizar2019approach}, whereas the new study (i.e., the third trial) was conducted in November of 2019. All trials were executed along two days. On the first day, the subjects answered the characterization form in order to allow dividing them into experimental groups. On the second day before the execution of the experiment, concepts of the security properties, OWASP high-level SRs, and defect types were reviewed by subjects in a training session. In the case of the new study, training on the inspection techniques was provided. After that, the inspection was conducted as follows.

All subjects had up to one hour to finish the review. In the original study, the control group used the OWASP high-level SRs as support during the review, while in the new study the control group used the PBR Black Hat approach. The experimental group used our approach. When the subjects finished the task, they had to fill out the follow-up questionnaire.

\section{Results} \label{sec:results}

In the following, we present the results of the trials conducted in the study. We first describe the results on defect detection effectiveness and efficiency. We also present more specific results on false positives introduced by inspectors and the types of defects identified by them. We end by evaluating the perception of the inspectors on the usefulness and ease of use of our approach. 

\subsection{Results on Defect Detection Effectiveness}
\label{subsec:effectiveness}

\begin{figure}[H]
    \centering    \includegraphics[width=0.9\textwidth]{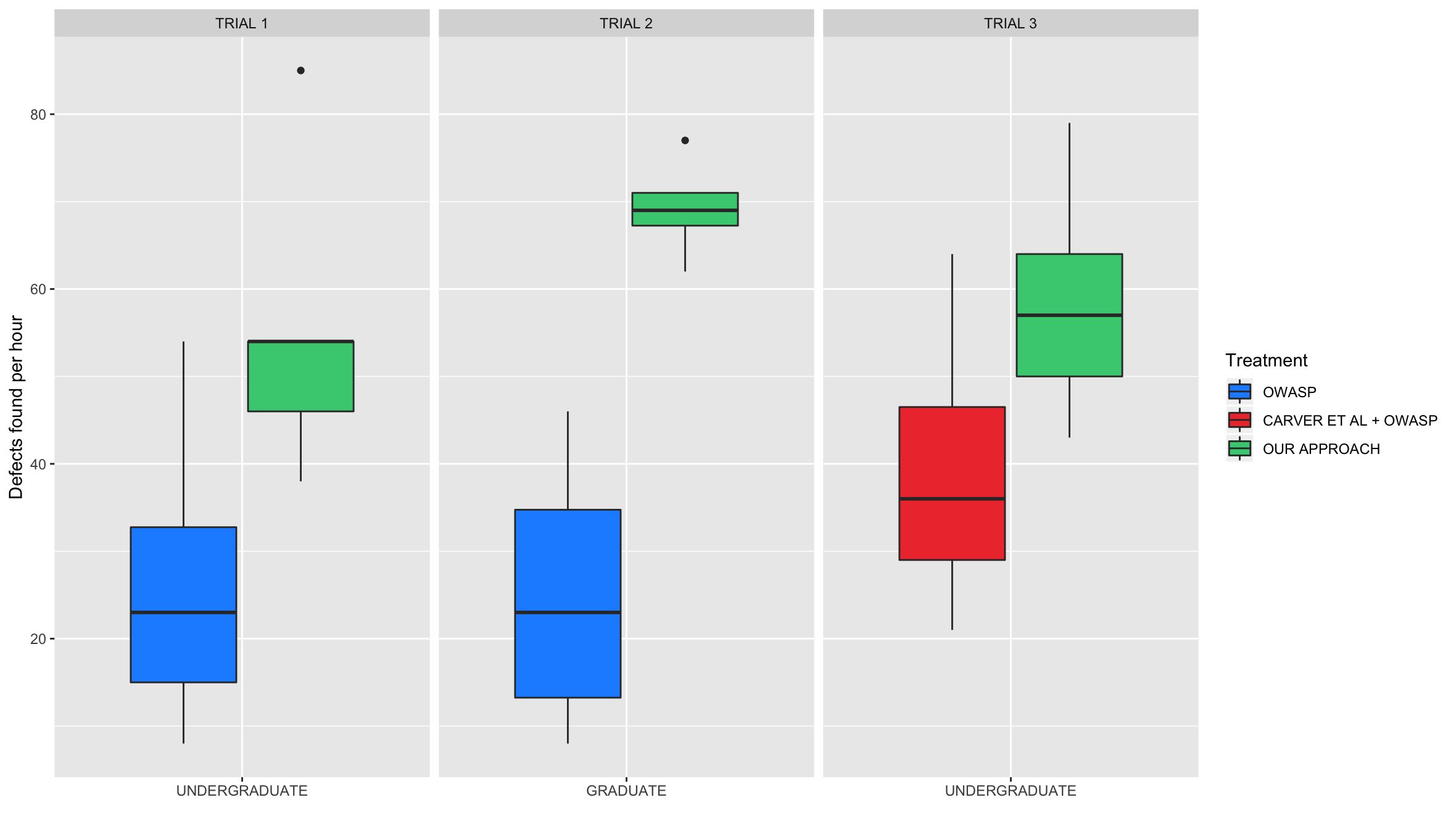}
    \caption{Defect detection effectiveness across the experiments}
    \label{fig:effectiveness_experiment}
\end{figure}

In this section we present the performance of inspectors in terms of effectiveness across the trials involved in the experiments. We analyze the number of defects found by inspectors who used our approach and other inspection methods with different levels of support. We wanted to understand the potential of our approach to detect security-related defects in agile requirements specifications of web applications. For this, we compared the performance of the effectiveness of our approach against the performance of the effectiveness of inspections conducted with the OWASP high-level SRs and the PBR Black Hat approach. Figure~\ref{fig:effectiveness_experiment} presents the defect detection effectiveness of each inspection technique by showing the distribution of the number of defects found by the subjects in each experimental trial. 

It is possible to observe that our approach was more effective than inspections conducted with the OWASP high-level SRs and the PBR Black Hat approach. In the original study (first and second trial), both experimental groups (green block) identified more defects than the control groups (blue block). For instance, in the first trial of the original study, conducted by undergraduate students, the experimental group (students who used our approach) identified in median, 54\% defects, while the control group (students who conducted the inspection by using the OWASP high-level SRs) identified 23\%. The difference was even higher when observing the performance of the second trial conducted by graduate students. Those who used our approach identified, in median, 69\% of the defects versus 23\% identified by the students who did not use it. This improvement can be explained for two reasons. (1) Often, graduate students have more experience than undergraduate students since the first ones have faced several real projects. We showed this trend in our characterization questionnaire. On the other hand, (2) in the second trial, we slightly modified the defect reporting form to ease its understanding and fulfillment. The reason was that in the first trial several inspectors mentioned that the defect reporting form was confusing. The change consisted of merging the columns A, IS and IR to understand better that those defect types do not have a 1 to 1 relationship with the security specifications such as the O column. In other words, we improved the design of the defect reporting form, while it remained to capture the same information.

Concerning the third trial (new study) conducted by undergraduate students, the effectiveness followed the same pattern as the original study. That is, students who used our approach identified more defects than students who used the PBR Black Hat approach. Note that in this trial, the control group also used the OWASP high-level SRs to support the PBR Black Hat approach. We decided to provide that support because the PBR Black Hat approach does not cover all the defect types introduced in the security specifications. It is noteworthy that the PBR Black Hat approach was not originally designed for the agile context, however, at the same time, its conception does not exclude this type of requirements. In the end, the experimental group (green block) identified, in median, 58\% of the defects, while the control group (red block) identified 38\%. 

When comparing the results between the new and the original study, we found that the PBR Black Hat approach (red block) improved the performance in terms of effectiveness when compared to the inspection conducted using only the OWASP high-level SRs (blue block). This indicates that the PBR Black Hat approach also helps in detecting security-related defects. This is outstanding if we consider that the PBR Black Hat approach was conceived in 2002 under different security concerns than those evaluated in our experiment.

We also wanted to test our null hypothesis on the effectiveness (H$_{01a}$ and H$_{01b}$), i.e., we checked whether the differences obtained in our experiments were significant to affirm or reject the hypotheses. To this end, we used the Mann-Whitney Test. In that sense, the results of the tests allowed to reject H$_{01a}$ and H$_{01b}$ for all experimental trials because we obtained p-values of 0.002, 0.012 and 0.004 for the first, second and third trial, respectively. This means there is a significant difference in terms of effectiveness when using our approach compared to conducting inspections with the OWASP high-level SRs and the PBR Black Hat approach. In other words, the amount of defects found of our approach can be considered significantly high compared to the other techniques involved in the experiments. Besides, to complement the p values results, we calculated Cohen's effect size which is a quantitative measure of the magnitude of a phenomenon. For all the experiments we obtained values of 3.46, 2.24 and 2.47 for the first, second and third trials, respectively. According to~\cite{vanvoorhis2007understanding}, this values can be considered as very large effect size. Thus, we can partially answer RQ1: Our approach has a positive impact on defect detection effectiveness with a very large effect size.

\subsection{Results on Defect Detection Efficiency}
\label{subsec:efficiency}

\begin{figure}[H]
    \centering
    \includegraphics[width=0.9\textwidth]{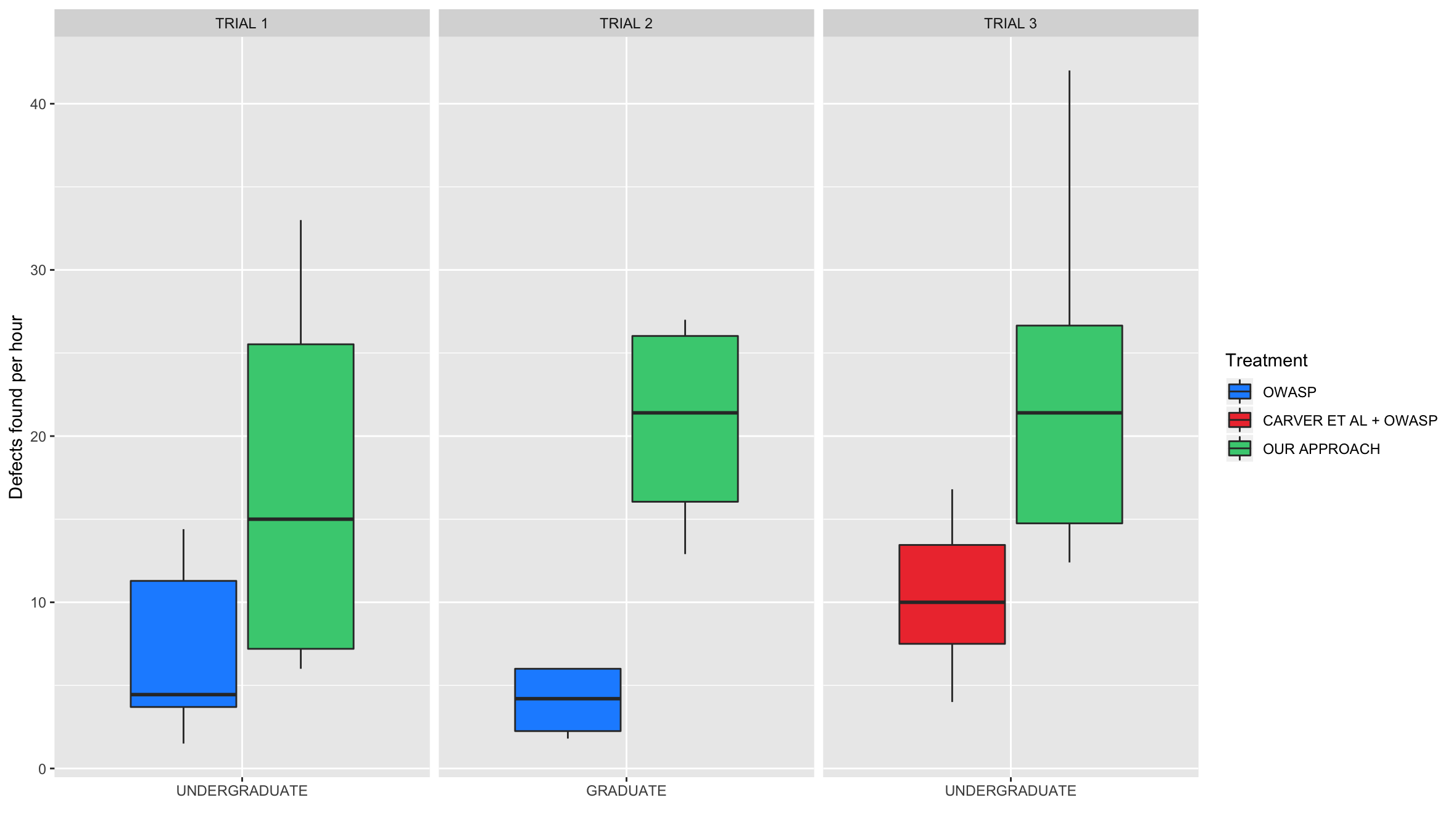}
    \caption{Defect detection efficiency across the experiments}
    \label{fig:effciency_experiment}
\end{figure}

After knowing the effectiveness of our approach across the different trials, we can question its efficiency by analyzing the defects found per hour by the inspectors. Figure~\ref{fig:effciency_experiment} shows the distribution of the efficiency of the subjects involved in the experiments.

Note that the efficiency follows the same pattern of the effectiveness, that is, the number of defects found per hour by the students who used our approach (green block) was higher than the students who used the OWASP high-level SRs and the PBR Black Hat approach (blue and red blocks, respectively).

In the original study, both experimental groups, in median, identified more defects per hour than the control groups. For instance, in the first trial conducted by the undergraduate students, our approach efficiency was 15 defects found per hour (we seeded 14 defects, but participants took less than one hour to complete their tasks). In contrast, the median of the undergraduate students who used only the OWASP high-level SRs was four. In the second trial conducted by the graduate students, the performance of our approach improved. The mean of our approach efficiency increased to 21 defects found per hour versus four defects found per hour by the graduate students who did not use it. This improvement is proportional to the effectiveness performance shown in Figure~\ref{fig:effectiveness_experiment}. Thus, we can explain these difference across the first two trials. 

In the third trial (new study) conducted by undergraduate students, the experimental group identified, in median, 22 defects per hour, whereas the control group identified, in median, ten defects. This means that the inspectors who used our approach identified defects faster than inspectors who used the OWASP high-level SRs and the PBR Black Hat approach. In this trial, we can see that undergraduate students who used our approach performed at the same level than graduate students (second trial) who used our approach. Under the same conditions, this would not be a typical performance considering the experience of the inspectors of each group. We believe that the training provided in the third trial on how our approach works and how the defect reporting form must be filled out may be a cause of this improvement.   

Regarding statistical hypothesis testing for efficiency, we found that the Mann-Whitney Test suggests rejecting our second null hypotheses (H$_{02a}$ and H$_{02b}$) with p-values of 0.02, 0.01 and 0.02 for the first, second and third trial, respectively. This means there is a significant difference in terms of efficiency when using our approach against the inspection conducted with the OWASP high-level SRs and the PBR Black Hat approach. In the same direction, we found that the relevance of this difference (Cohen's effect size~\cite{vanvoorhis2007understanding}) was large for all the experiments (1.56, 3.29 and 2.32 for the first, second and third trials, respectively). With this information, we can fully answer RQ1. Our approach has a positive impact on security defect detection effectiveness and efficiency when compared to directly using the OWASP high-level SRs and to using the PBR Black Hat approach.

\subsection{Results on False Positives}

\begin{figure}[H]
    \centering
    \includegraphics[width=0.8\textwidth]{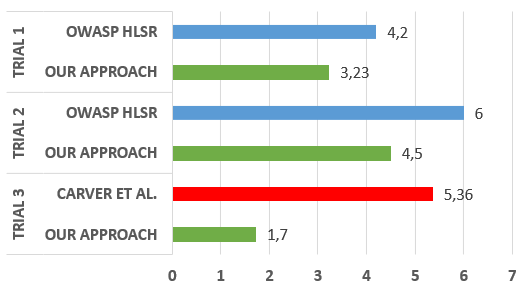}
    \caption{Average of false positives by trial and technique}
    \label{fig:false_positives}
\end{figure}

To understand to which extend our approach leads inspectors to report defects that are not truly defects (false positives), we analyzed the output of each inspector of the trials to manually and in pairs classify whether the reported defect concerns a true defect or a false positive. False positives may affect both efficiency and effectiveness, thus these results provide additional insights to answer RQ1. Note that this analysis was conducted in pairs of researchers. We present Figure~\ref{fig:false_positives} to provide an overview of the false positives by trial and groups involved in the experiments.

According to Table~\ref{fig:false_positives}, we see that all the trials followed the same pattern. Inspectors who used our approach introduced less false positives than inspectors who conducted the inspection by using the OWASP high-level SRs and the PBR Black Hat approach. For instance, in the second trial our approach led the inspectors to introduce, in average, 4.5 false positives while the inspectors who used only the OWASP high-level SRs introduced in average 6.0. In the third trial, where we evaluated our approach against the PBR Black Hat approach, we see that our approach led the inspectors to introduce, in average, 1.7 false positives against 5.4 introduced by the inspectors who used the PBR Black Hat approach. When investigating the possible causes of these findings, we believe that the context factors involved in each experiment, and which were presented in Table~\ref{tab:context_experiment}, influenced to improve the results of the third trial with respect to the first and second one. For instance, we provided additional training in the new study by teaching the inspectors to follow the task description and our reading technique. Overall, this can be seen as a benefit of our approach since this factor is closely tied to software quality assurance. In other words, if we link these results on false positives with the results on defect detection effectiveness and efficiency, when using our approach inspectors tend to find more defects in less time, making less mistakes. Therefore, these results support again our findings on effectiveness and efficiency presented before. 

\subsection{Analysis Across Experimental Trials}
\label{subsec:results_across_studies}

We present a synthesis of our findings from the first, second and third trial conducted in this study as follows. We provide overall median scores for effectiveness and efficiency performance in Table~\ref{tab:overall_scores_ordered}. In order to provide a high-level overview of our findings, we rank the results by efficiency performance. We consider efficiency instead of effectiveness given the importance of time spent on agile methods. We also present the subject level, undergraduate (U) or graduate (G), the treatments and the experimental trial number. 

\begin{table}[H]
\caption{Overall mean scores of performance sorted by efficiency}
\label{tab:overall_scores_ordered}
\begin{tabular}{c p{1.6cm} c c c c}
\hline\noalign{\smallskip}
Ranking & Efficiency (def/hour) & Effectiveness & Level & Treatment & Trial
\\\noalign{\smallskip}\hline\noalign{\smallskip}

1 & \multicolumn{1}{c}{22} & 58,4 & U & Our Approach & 3 \\
2 & \multicolumn{1}{c}{21} & 68,9 & G & Our Approach & 2 \\
3 & \multicolumn{1}{c}{15} & 54,2 & U & Our Approach & 1 \\
4 & \multicolumn{1}{c}{10} & 38,3 & U & PBR Black Hat approach & 3 \\
5 & \multicolumn{1}{c}{4}  & 23,6 & G & OWASP high-level SRs & 2  \\
6 & \multicolumn{1}{c}{4}  & 23,1 & U & OWASP high-level SRs & 1
\\\noalign{\smallskip}\hline
\\\multicolumn{5}{l}{%
\begin{minipage}{7cm}%
\end{minipage}%
}
\end{tabular}
\end{table}

If we compare the performance across the trials we see that the best three performances were obtained by the inspectors who used our approach. In other words, experimental group always performed better than control group. Taking a look at the performance by educational level (undergraduate and graduate), we see that this factor had limited influence on the performance of the inspectors. In our experiments, performance was mainly defined by the treatment given to inspectors, that is, if a support such as structured technique is provided, the performance of the inspectors, regardless of his/her education level, can improve. This is supported by analyzing the performance of the inspectors who used the PBR Black Hat approach. These inspectors performed better than inspectors who did not received support in form of a structured technique.

\subsection{Results on Types of Defects Identified}

We also wanted to determine to what extent our approach helps to identify incomplete, inconsistent, incorrect and ambiguous security-related aspects. Table~\ref{tab:summary_experiment} shows the distribution of the defects found per type by each participant in all the trials. We highlighted the discarded subjects as mentioned in Section~\ref{subsec_experimentContext}. For a better understanding of the data shown in Table~\ref{tab:summary_experiment}, we summarize in Table~\ref{tab:summary_defects_per_type} the average percentage of defects found by defect type according to the techniques involved in the experiments. 

We observed that inspectors who used our approach on average identified 92,4\% of omission defects. I.e., almost all such defects were identified. Regarding the other defect types, we observed that these inspectors found 45,2\% of ambiguity defects, 60,5\% of inconsistency defects, and 42,6\% of defects related to incorrect facts. In total, inspectors who used our approach in average identified 60\% of the seeded defects. In comparison with the performance of the inspectors who used the PBR Black Hat, they did not perform at the same level as inspectors who used our approach. For instance, inspectors who used the PBR Black Hat approach found in average 32,2\% of omission defects, 29,5\% of ambiguity defects, 54,8\% of inconsistent defects and 31\% of incorrect facts defects. Considering all defect types, in average, this group found 40\% of the defects against 60\% found by subjects using our approach. The landscape is even poorer if we compare the performance of the inspectors who conducted the inspection by using only the OWASP high-level SRs, which in average found only 23\%.

Given these results, we are confident that our approach helps identifying omission defects, that is, to detect security-related aspects that were not considered or were not covered by the security specifications originally created by requirements analysts. We also consider that our approach contributes identifying inconsistency defects since more than half of these defects were identified. Comparing the other defect types, we see that our approach is slightly better than the PBR Black Hat approach and the inspection conducted with only the OWASP high-level SRs.

\begin{table}[H]
\centering
\footnotesize
\caption{Detail of the inspectors' performance in all trials}
\label{tab:summary_experiment}
\begin{tabular}{c c c c c c c c c c c c c c}
\hline\noalign{\smallskip}
ID  & Trial & Treatment & Time & OM & \% & AM & \%  & IS & \%  & IF & \%  & $\sum$  
\\\noalign{\smallskip}\hline\noalign{\smallskip}
1&1&Our approach& 00:55  &4&100&2&50&1&25&0&0&7\\
2&1&Our approach& 01:00  &4&100&1&25&1&25&0&0&6\\
3&1&Our approach& 00:48  &4&100&0&0&0&0&0&0&4\\
4&1&OWASP HLSR& 00:50 &3&75&1&25&0&0&0&0&4\\
5&1&OWASP HLSR& 00:38 &3&75&1&25&3&100&0&0&7\\
6&1&OWASP HLSR& 00:44 &2&50&1&25&0&0&0&0&3\\
7&1&Our approach&  00:40 &3&75&1&25&2&50&0&0&6\\
\cellcolor[HTML]{C0C0C0}8&\cellcolor[HTML]{C0C0C0}1&\cellcolor[HTML]{C0C0C0}OWASP HLSR& \cellcolor[HTML]{C0C0C0}00:40 &\cellcolor[HTML]{C0C0C0}0&\cellcolor[HTML]{C0C0C0}0&\cellcolor[HTML]{C0C0C0}0&\cellcolor[HTML]{C0C0C0}0&\cellcolor[HTML]{C0C0C0}0&\cellcolor[HTML]{C0C0C0}0&\cellcolor[HTML]{C0C0C0}1&\cellcolor[HTML]{C0C0C0}50&\cellcolor[HTML]{C0C0C0}1\\
9&1&OWASP HLSR& 00:30  &0&0&1&25&0&0&1&50&2\\
10&1&Our approach& 00:50 &4&100&0&0&0&0&1&50&5\\
11&1&Our approach& 00:20 &4&100&0&0&0&0&0&0&4\\
12&1&OWASP HLSR&  00:35  &0&0&1&25&1&25&0&0&2\\
13&1&Our approach& 00:36 &3&75&0&0&0&0&0&0&3\\
14&1&OWASP HLSR&  00:31  &0&0&0&0&2&50&0&0&2\\
\cellcolor[HTML]{C0C0C0}15&\cellcolor[HTML]{C0C0C0}1&\cellcolor[HTML]{C0C0C0}OWASP HLSR&  \cellcolor[HTML]{C0C0C0}00:28  &\cellcolor[HTML]{C0C0C0}0&\cellcolor[HTML]{C0C0C0}0&\cellcolor[HTML]{C0C0C0}0&\cellcolor[HTML]{C0C0C0}0&\cellcolor[HTML]{C0C0C0}1&\cellcolor[HTML]{C0C0C0}25&\cellcolor[HTML]{C0C0C0}0&\cellcolor[HTML]{C0C0C0}0&\cellcolor[HTML]{C0C0C0}1\\
16&1&Our approach& 00:20 &4&100&3&75&2&50&2&100&11\\
17&1&OWASP HLSR&  00:25  &1&25&2&50&2&50&0&0&5\\
18&1&Our approach& 00:26 &2&50&0&0&0&0&0&0&2\\
19&1&Our approach& 00:17 &4&100&1&25&2&50&&0&7\\
20&1&OWASP HLSR&  00:25  &0&0&1&25&2&50&0&0&3\\
21&1&OWASP HLSR&  00:25  &0&0&2&50&3&100&1&50&6\\
22&1&Our approach& 00:20 &4&100&0&0&0&0&0&0&4\\
23&1&Our approach& 00:20 &4&100&1&25&2&50&0&0&7\\
24&1&Our approach& 00:15 &4&100&1&25&2&50&0&0&7\\
25&1&OWASP HLSR&  00:15  &2&50&0&0&1&25&0&0&3\\

\cellcolor[HTML]{C0C0C0}26&\cellcolor[HTML]{C0C0C0}2&\cellcolor[HTML]{C0C0C0}OWASP HLSR&  \cellcolor[HTML]{C0C0C0}00:33  &\cellcolor[HTML]{C0C0C0}0&\cellcolor[HTML]{C0C0C0}0&\cellcolor[HTML]{C0C0C0}0&\cellcolor[HTML]{C0C0C0}0&\cellcolor[HTML]{C0C0C0}1&\cellcolor[HTML]{C0C0C0}25&\cellcolor[HTML]{C0C0C0}0&\cellcolor[HTML]{C0C0C0}0&\cellcolor[HTML]{C0C0C0}1\\
27&2&Our approach& 00:35 &4&100&2&50&2&50&2&100&10\\
28&2&Our approach& 00:21 &4&100&3&75&2&50&0&0&9\\
29&2&Our approach& 00:20 &4&100&3&75&2&50&0&0&9\\
30&2&OWASP HLSR&  00:50  &0&0&0&0&2&50&0&0&2\\
31&2&Our approach& 00:37 &4&100&1&25&2&50&1&50&8\\
32&2&OWASP HLSR&  01:00  &0&0&3&75&2&50&1&50&6\\
33&2&OWASP HLSR&  00:40  &1&25&1&25&1&25&1&50&4\\

34&3&Our approach& 00:10 &2&50&2&50&2&50&1&50&7\\
35&3&Our approach& 00:15 &4&100&2&50&3&75&0&0&9\\
36&3&PBR Black Hat& 00:30 &1&25&1&25&2&50&1&50&5\\
37&3&PBR Black Hat& 00:24 &0&0&1&25&2&50&1&50&4\\
38&3&Our approach& 00:16 &3&75&1&25&3&75&1&50&8\\
39&3&Our approach& 00:18 &3&75&1&25&2&50&1&50&7\\
40&3&PBR Black Hat& 00:25 &1&25&3&75&3&75&0&0&7\\
41&3&PBR Black Hat& 00:26 &0&0&0&0&3&75&1&50&4\\
42&3&Our approach& 00:28 &4&100&2&50&3&75&1&50&10\\
43&3&Our approach& 00:26 &2&50&2&50&2&50&0&0&6\\
44&3&PBR Black Hat& 00:30 &2&50&2&50&2&50&1&50&7\\
45&3&PBR Black Hat& 00:28 &4&100&0&0&2&50&0&0&6\\
46&3&Our approach& 00:25 &4&100&2&50&2&50&1&50&9\\
\cellcolor[HTML]{C0C0C0}47&\cellcolor[HTML]{C0C0C0}3&\cellcolor[HTML]{C0C0C0}Our approach& \cellcolor[HTML]{C0C0C0}00:30 &\cellcolor[HTML]{C0C0C0}1&\cellcolor[HTML]{C0C0C0}25&\cellcolor[HTML]{C0C0C0}0&\cellcolor[HTML]{C0C0C0}0&\cellcolor[HTML]{C0C0C0}0&\cellcolor[HTML]{C0C0C0}0&\cellcolor[HTML]{C0C0C0}0&\cellcolor[HTML]{C0C0C0}0&\cellcolor[HTML]{C0C0C0}1\\
48&3&Our approach& 00:30 &2&50&1&25&3&75&1&50&7\\
49&3&Our approach& 00:30 &3&75&1&25&2&50&2&100&8\\
50&3&Our approach& 00:31 &2&50&1&25&3&75&2&100&8\\
51&3&PBR Black Hat& 00:36 &1&25&0&0&2&50&1&50&4\\
52&3&PBR Black Hat& 00:36 &4&100&2&50&2&50&1&50&9\\
53&3&PBR Black Hat& 00:43 &0&0&2&50&3&75&1&50&6\\
54&3&PBR Black Hat& 00:45 &0&0&0&0&3&75&0&0&3\\
55&3&PBR Black Hat& 00:40 &1&25&2&50&1&25&0&0&4\\
56&3&Our approach& 00:53 &4&100&3&75&3&75&1&50&11

\\\noalign{\smallskip}\hline
\\\multicolumn{7}{l}{%
\begin{minipage}{7cm}%
OM: Omission\hspace{0.3cm}AM: Ambiguity\hspace{0.3cm}IS: Inconsistency\hspace{0.3cm}IF: Incorrect Fact%
\end{minipage}%
}
\end{tabular}
\end{table}

\begin{table}[H]
\caption{Distribution of defects found per treatment}
\label{tab:summary_defects_per_type}
\begin{tabular}{p{3cm} c c c c c}
\hline\noalign{\smallskip}
Treatment                             & \multicolumn{1}{l}{Omission} & \multicolumn{1}{l}{Ambiguity} & \multicolumn{1}{l}{Inconsistency} & \multicolumn{1}{l}{Incorrect Fact} & \multicolumn{1}{l}{Total}
\\\noalign{\smallskip}\hline\noalign{\smallskip}
OWASP high-level SRs                          & 23,2\%                                   & 25,8\%                                    & 36,5\%    & 14,8\%    & 23\% \\
PBR Black Hat + OWASP   & 32,8\%    & 29,5\%   & 54,8\%  & 31,8\%   & 40\% \\                        
Our Approach   & 92,4\%    & 45,2\%   & 60,5\%   & 42,6\%  & 60\%                                  \\\noalign{\smallskip}\hline
\end{tabular}
\end{table}

Note that the results of the ambiguity defect are not very different among the experiments, that is, neither the support of our technique nor the support of the PBR Black Hat approach generates a relevant improvement compared to an ad-hoc inspection (using only OWASP high-level SRs).

Despite promising results in identifying defects of omission and inconsistency, we believe that the verification questions that are part of our approach should be reviewed to improve its effectiveness. This is specially important for ambiguity and incorrect fact defects. It is also particularly interesting that the percentage to identify incorrect fact defects was low in all the inspections. Probably, this happened due to the additional domain knowledge that is usually needed to detect such problems. This may indicate the difficulty of identifying this type of defects is higher than the others. In summary, we must consider improvements to increase the effectiveness in detecting this kind of defects.

\subsection{Perception of the Inspectors on the Usefulness and Ease of Use}

\begin{figure}[H]
    \centering
    \includegraphics[width=0.9\textwidth]{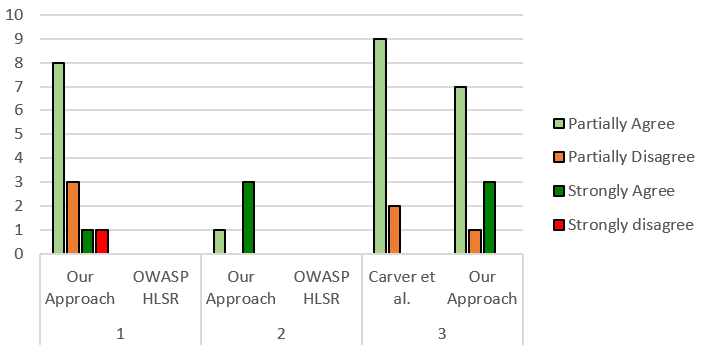}
    \caption{Perception of the inspectors to the statement ``I found the approach easy to use"}
    \label{fig:ease_to_use}
\end{figure}

\begin{figure}[H]
    \centering
    \includegraphics[width=0.9\textwidth]{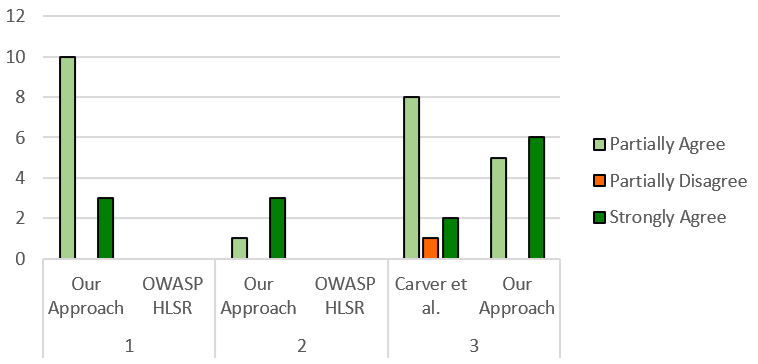}
    \caption{Perception of the inspectors to the statement ``Using the approach improved my performance (find defects faster)"}
    \label{fig:usefulness}
\end{figure}

After inspectors conducted the inspection task by reviewing the security specifications provided in the experiments, we asked whether they found our approach useful and easy to use. Through the TAM questionnaire~\cite{davis1989perceived}, we wanted to know about their perceptions on using our approach and the PBR Black Hat approach. Figure~\ref{fig:ease_to_use} and Figure~\ref{fig:usefulness} show the frequencies of the responses of the inspectors that measures their perception on ease of use and usefulness, respectively. These figures present green tones that indicate agreement and red tones that indicate disagreement with our statements of ease of use and usefulness. Note that dark tones indicate a strong agreement/disagreement and light tones indicate a partial agreement/disagreement. 

We start by analyzing the responses of the inspectors that measures their perception related to the ease of use of the reading-based approaches. Figure~\ref{fig:ease_to_use} shows the perception of the inspectors for the statement \say{I found the approach easy to use}.

First, we can see that the we have a higher concentration of red tones in the first trial. In this case,

the perception of inspectors about the ease of use improved across the trials. For instance, in the first trial, 4 out of 13 (31\%) strongly or partially disagreed with the statement \say{I found the approach easy to use}. Only one inspector strongly agreed with. In the second and third trial, the picture was different. Just one out of 15 (6\%) inspectors who used our approach in these trials partially disagreed. The other 14 inspectors partially (8) and strongly (6) agreed. A large number of disagreements and partial agreements in the first trial indicated that improvements should be added to facilitate the use of the approach and then improve the detection of defects. In that direction, the inspectors proposed some points that could enhance the ease of use of the approach, such as providing a lighter document, modifying the design of the defect reporting form and showing an example of how to fill it out correctly. This feedback was taken into account to perform the second and third trials. In fact, as mentioned in Section~\ref{subsec:effectiveness}, the inspectors' defect detection effectiveness improved, in our opinion, because we introduced some changes to the defect reporting form such as merging the columns related to ambiguous, inconsistency and incorrect facts, and simplifying the number of steps in the task description document. 

When comparing the perception of ease of use of the approaches involved in the third trial (our approach against the PBR Black Hat approach), we see that, in principle, inspectors found our approach easier to use. For instance, nine of the inspector who used the PBR Black Hat approach partially agreed and none of them strongly agreed with the sentence about ease of use. This indicates they faced difficulties to follow the inspection under its guidance. In contrast, seven inspectors who used our approach partially agreed and three strongly that the approach was easy to use.   

We were also interested in knowing the perception of the inspectors in relation to the usefulness of the approach. We defined usefulness as a metric of productivity based on efficiency. More specifically, we asked whether using the approach would improve their performance when conducting the security requirements inspection (find defects faster). Table~\ref{fig:usefulness} gives an insight into the perception of the inspectors related to the usefulness of our approach.

In this case, the perception of the inspectors regarding the usefulness of our approach in the trials was, in general, positive. For instance, all the inspectors who used our approach (28) partially or strongly agreed with the sentence we provided about usefulness. More specifically, 12 out of them (43\%) strongly agreed and 16 (57\%) partially agreed. However, note that most of the partial agreements were reported in the first trial (10 out of 16). This perception arises from the difficulties faced by the inspectors in that trial. For example, one inspector stated the following: \say{The review may be exhausting and time-consuming because the task description document is not lightweight}. Another one stated: \say{It would be better to automate the proposal}. The inspectors also mentioned some difficulties faced, such as \say{The verification questions could indicate better how to identify defects such as ambiguity or incorrect facts}. We believe these barriers may affect the performance of the inspectors and, thus, the perception of usefulness. 

Concerning the perception of the inspectors who used the PBR Black Hat approach in the third trial, we observed that most of the inspectors found it useful. In this case, just one inspector partially disagreed with the sentence about the usefulness we provided. 8 out of 11 (73\%) partially agreed and 2 out of 11 (18\%) strongly agreed. This reinforces the idea that novice inspectors need support to review security-related aspects. Therefore, approaches addressing these concerns help to identify defects that would not be easily identified by ad-hoc inspections. 

Nevertheless, if we compare the perception of usefulness between our approach and the PBR Black Hat approach, we see that more inspectors strongly agreed with our approach. This reflects that inspectors felt more comfortable using our approach. Also, this perception is aligned with the performance of inspectors across the trials, since inspectors who used our approach performed better than inspectors who did not use it. 

In summary, to the question of how do the inspectors perceive the usefulness and ease of use of our approach, we have that inspectors who used our approach found it, in principle, easy to use and usefulness. Indeed, the perception of the inspectors was positive in comparison with other security inspection techniques. However, we have several challenges to improve these aspects, such as automating our approach and improving the security verification questions. 

\section{Threats to Validity} \label{sec:threatsValidity}

We report several threats to validity following the recommendations by Wohlin et al~\cite{wohlin2012experimentation} that we considered or mitigated during the design and execution of the original and new study.

\subsection{Internal validity}
\label{subsec:internal}

First, aiming to avoid personal bias, we used researcher triangulation to collect and analyze all data. The profile of researchers varies in relation to experience but remains in relation to the application domain, in this case, software engineering. One master and one Ph.D student in informatics with the supervision of one senior researcher were involved in this research triangulation. We carefully reviewed the extraction of defects and calculation of the percentages of performance of the inspectors. 

Second, we characterized all the subjects with the aim of removing confounding factors. The characterization allowed us to apply the blocking principle by distributing the participants so that these characteristics were equally distributed among the experimental and control groups. Table~\ref{tab:characterization} highlights the subjects who were chosen because of the partition principle. Random assignment was employed for subjects with similar characteristics. 

We also consider that the performance of participants could be affected if inspectors try to guess the purpose of the experiment. In the original study, participants were not aware of the existence of experimental versus control groups or whether they belonged to different groups. Participants were told only that they were supposed to perform the task of detecting security-related defects based on a given requirements specifications. Thus single blinding was used to minimize biases. Regarding the new study, participants were aware
of the group treatment because we introduce them both the inspection techniques involved in the experiment. We examined the responses by the groups to see if they resembled closely with treatment responses. However, we did not find any evidence of treatment diffusion across the groups.

Finally, regarding training, we provided the same examples of user stories, defects, security specifications and controls to the experimental and control group of all the trials, so any potential bias is similar for all the subjects. In summary, participants received the same training. 

\subsection{Construct validity}
\label{subsec:construct}

In our evaluation, we analyzed the suitability of our approach in terms of the number of defects detected, defect types and number of false positives. To this end, quantitative analysis was performed. We used metrics such as effectiveness and efficiency that are commonly used in inspection studies that are empirically evaluated. We also analyzed the perceptions on usefulness and ease of use of the inspectors when using our approach and the PBR Black Hat approach. To this evaluation, qualitative analysis was considered. We used the TAM questionnaire~\cite{davis1989perceived}, which has also been widely used and evaluated~\cite{turner2010does} to measure the acceptances of techniques, applications and technologies.

\subsection{Conclusion validity}
\label{subsec:conclusion}

Reliability of measures is an important consideration to draw valid conclusions about the results. We used the Mann-Whitney Test with the aim of determining  whether we reject or not our null hypotheses. The decision of using this method was supported according to the distribution of our independent samples, in this case, the treatments of the experiments. Besides statistical significance, we used the Cohen's \textit{h} metric that is a measure of distance between two proportions or probabilities. With this, we determine whether the difference of our results can be considered as small, medium, or large. In our study, we found the relevance scores of the experimental group was significantly better than the control group. 

Regarding number of participants, we had 56 participants divided into groups that characterize different samples. Factors such as type of study (original and new), position of the students (undergraduate and graduate) and type of treatment (experimental and control) were considered among the experiments.

\subsection{External validity}
\label{subsec:external}

To mitigate this kind of validity threat, the sample population should be representative of the population we want to evaluate. Regarding the subject representativeness, we used students to represent novice inspectors. Using students as subjects remains a valid simplification of real-life settings needed in laboratory contexts~\cite{falessi2018empirical}. In the studies, participants are representative of students in computer science enrolled in two different graduate and undergraduate courses. For all of them, we provided concepts related to security principles and inspection techniques. Regarding the objects, we created the agile specifications following the quality guidelines proposed by Lucassen~\cite{lucassen2015forging}. We also peer-reviewed the requirements specifications and the seeded defects in terms of their representativeness. Through the new study, we have demonstrated the applicability of the requirements specifications created in identifying security-related defects. As we planned to conduct a limited amount of trials with a limited amount of subjects, the experiment package is available for external replications.

\section{Discussion} \label{sec:discussion}

This work brought up several further questions that have strong implications on future research. Therefore, in the following, we discuss several of these aspects in more depth.

\subsection{Suitability of the OWASP high-level SRs}
\label{subsec:suitability}

We are aware that not all OWASP high-level SRs might be useful in all situations. However, we are confident that as a starting point it is useful to have a basis that allows novice inspectors, at least, to consider the basic needs to deal with security.  

\subsection{Generalization of our approach}
\label{subsec:generalization}

We know that not only security is challenging in agile projects. Indeed, it seems that other NFRs such as maintainability and performance are often ignored or ill-defined in this context. Moreover, plan-driven software projects may face similar problems. This provides an opportunity to extend our approach, e.g., considering other types of inputs such as open textual requirements and covering other quality characteristics such as portability and usability. Currently, knowledge on the available verification techniques to assure these quality characteristics are met is scattered and limited. Furthermore, those techniques are commonly not properly integrated into the agile development philosophy. Thus, we consider this, a first step in this direction was conducted by investigating security, a specific product quality characteristic. 

\subsection{Implications of our research for practitioners}
\label{subsec:implications_practitioners}

Unfortunately, security problems tend to be postponed to later stages ~\cite{sampaio2016earlydetection}. For practitioners, our approach provides a way to detect defects related to security aspects that should be specified. According to the results of the original and new study, we are confident to say that in principle our approach supports novice inspectors by providing guidance in applying a reading technique that will help them to identify defects in agile requirements specifications of web applications. In addition, we designed the approach in such a way that it works without expensive review cycles, aligned with the agile philosophy. We see three main potential benefits of this approach. First, narrowing the security knowledge gap that exists between experts and novice inspectors. Second, the reading technique provides a strong focus on security aspects. In this way, the team can avoid discussing obvious issues and focus on important, difficult, security-specific aspects of the review. Third, we saw that our approach provided positive results regarding the performance of individual inspectors when conducting the security inspection.

\subsection{Implications of our research for researchers}
\label{subsec:implications_researches}

This work contributes already to closing an important literature gap that exists with respect to security requirements verification in the agile context. Based on the scarce literature on the topic, we believe that our approach constitutes an interesting starting point to discuss in depth verification activities centered on security in agile contexts. For us, the results strengthen our confidence in further extending our approach to scale its usability up to practical settings covering a full, tool-supported process integration, which was not (and could not be) in scope of a development in our research-centric environment.

\subsection{Limitations} \label{sec:limitations}

We concentrated on a set of specific security properties and high-level SRs from the OWASP (matching security sub-characteristics also described in the SQuaRE quality model). There are several security standards that are different from the ones provided by OWASP. Thus, we could complement the security vision of our approach with other standards.

Moreover, given the complexity of working with NLP in RE, there is a limitation related to the completeness of the keyword repository needed to link the user stories with the security properties. To deal with this, we decided to consider synonyms regarding the initial set of keywords.

We only evaluated our approach with a use case scenario (two user stories with their security specifications) as a starting point for detecting defects related to security. However, additional use cases can also be provided as input to the participants and may give us insights on the suitability of our approach in different contexts.
We are currently designing an industrial case study to evaluate the coverage of detecting security defects by our approach for a real software system. The results will provide evidence on how the approach generalizes when applied with the help of security analysts without the time and other experimental constraints.

We are also aware that the security specifications involved in the experiments constitute a limitation of the study. In a perfect scenario, we would have security concerns specified by companies or independent practitioners, but often this information is restricted. Therefore, we invested our best efforts to carefully create and verify the specifications on their representativeness. Nevertheless, external replications, including a wider range of user stories and security specifications, are needed to improve external validity of our results.

\section{Concluding Remarks} \label{sec:concluding_remarks}

This work addresses a gap in the literature concerning the absence of verification techniques for security in agile requirements engineering and its lack of empirical evidence. It is well-known that the poor definition of NFRs, minimum documentation and lack of requirements verification are among the most important concerns of software requirements engineering researchers~\cite{chung2012non},~\cite{nuseibeh2000requirements}. Therefore, we presented an approach for reviewing security-related aspects in agile requirements specifications of web applications, which we empirically evaluated via three controlled experiment trials. In the following, we summarize our conclusions and we discuss potential practical implications of our research.

The three trials concerned evaluating the effectiveness, efficiency, usefulness and ease of use of our approach when compared to an ad-hoc inspection supported with the OWASP high-level SRs and another one supported with the PBR Black Hat approach. In the combined analysis of all the controlled experiments, participants in the experimental group performed significantly better than participants in the control group in terms of effectiveness and efficiency, i.e., participants who used our approach identified more defects in less time than participants who did not use it. We also identified that participants who used our approach found it, in principle, useful and easy to use.  

Future work includes evaluating the performance of using our approach in industrial settings. We want to reach out to better understand the performance of using our approach in real settings, as well as further information to help us better addressing practitioners needs. Therefore, we might have to provide tool support for the application of our approach, in such a way that, for instance, applying the reading technique could be guided by the FESRAS framework.

\bibliographystyle{spmpsci}      
\bibliography{sigproc}   

\end{document}